\documentclass[a4paper,11pt]{article}
\usepackage{cite}
\usepackage{fullpage,setspace,hyperref,comment,caption}
\usepackage[font=footnotesize,labelfont=bf,margin=1cm]{caption}
\usepackage{cite} 
\usepackage{color}
\usepackage{amsfonts}
\usepackage{amsmath}
\usepackage{amssymb}
\usepackage{todonotes} 
\usepackage[utf8]{inputenc}
\usepackage{slashed}
\usepackage{graphicx}
\usepackage{pstricks-add}
\usepackage{tikz}
\usetikzlibrary{arrows,snakes,backgrounds}

\newcommand{\diag}{\mathrm{diag}}
\newcommand{\be}{\begin{equation}}
\newcommand{\ee}{\end{equation}}

\def\Lax{{\mathfrak L}}

\newcommand \arxivlink [1]{\href{http://arxiv.org/abs/#1}{\tt arXiv:#1}}
 
\begin{document}
\begin{titlepage}
\vfill

\begin{flushright}
\end{flushright}

\vfill

\begin{center}
   \baselineskip=16pt
   	{\Large \bf Integrability and non-Integrability in ${\mathcal N}=2$ SCFTs and their Holographic Backgrounds}
   	\vskip 2cm
 	{\large \bf Carlos Nunez$^a$\footnote{\tt C.Nunez@Swansea.ac.uk}, Dibakar Roychowdhury$^a$\footnote{\tt dibakarphys@gmail.com} and Daniel C. Thompson$^{a}$\footnote{\tt d.c.thompson@swansea.ac.uk}}
   	\vskip .6cm
   	{\it  $^a$ Department of Physics, Swansea University, \\ Swansea SA2 8PP, United Kingdom \\ \ \\}
   	\vskip 2cm
\end{center}

\begin{abstract}
We show that the string worldsheet theory of Gaiotto-Maldacena holographic  duals to ${\mathcal N}=2$ superconformal field theories generically fails to be classically integrable.  We demonstrate numerically that the dynamics of a winding string configuration possesses a non-vanishing Lyapunov exponent. Furthermore an analytic study of the Normal Variational Equation fails to yield a Liouvillian solution.   An exception to the generic non-integrability of such backgrounds is provided by the non-Abelian T-dual of $AdS_5\times S^5$; here by virtue of the canonical transformation nature of the T-duality classical integrability is known to be present. 
\end{abstract}

\vfill

\setcounter{footnote}{0}
\end{titlepage}
\tableofcontents 

\section{Introduction}
 One of the most attractive features of the correspondence between ${\cal N}= 4$ supersymmetric Yang-Mills theory and type IIB superstrings in $AdS_5 \times S^5$ \cite{Maldacena:1997re}, is the presence of {\em integrability} in the planar limit \cite{Beisert:2010jr}. This triumph begs the question to what extent can integrable structures be found in less symmetric gauge theories.   
In this note we will concern ourselves with the study of integrable structures--or lack thereof--in the worldsheet theories that describe, via holography, quantum field theories.

For the $AdS_5 \times S^5$ superstring classical integrability is ensured since the Lagrangian equations of motion can be expressed as the flatness of Lax connection \cite{Bena:2003wd}.  A similar Lax formulation of the dynamics is available for strings propagating in the gravity duals to  marginal Leigh-Strassler  real  $\beta$-deformation that preserve ${\cal N}=1$ supersymmetry \cite{Leigh:1995ep}.  
However, when the $\beta$-deformation is complex,  integrability is not present \cite{Roiban:2003dw,Frolov:2005ty,Berenstein:2004ys,Giataganas:2013dha}.     

In the same line, the   dual to the ${\cal N}=1$ superconformal Klebanov-Witten theory is given by strings in $AdS_5 \times T^{1,1}$ \cite{Klebanov:1998hh}.  Despite  the preserved supersymmetry,  the large non-abelian isometry group and geodesic integrability of $T^{1,1}$, it has been shown that certain classical string configurations are chaotic and hence integrability is not present in the corresponding gauge theory \cite{Basu:2011di}.    Conversely the comparatively recent  development of so-called $\lambda$-deformed \cite{Sfetsos:2013wia,Hollowood:2014qma} and $\eta$-deformed \cite{Klimcik:2002zj,Delduc:2013qra} theories  provide backgrounds as developed in \cite{Sfetsos:2014cea,Arutyunov:2015mqj,Borsato:2016zcf,Chervonyi:2016ajp,Borsato:2016ose} with fewer/no isometries and no supersymmetries  but yet are integrable; such theories are thought to  be described  by  quantum group deformations of  the $AdS_5 \times S^5$ superstring.\footnote{To be precise the $\eta$-deformed backgrounds satisfy conditions of global scale invariance that are weaker  than full local conformal invariance and hence obey field equations of a modification to type II supergravity \cite{Arutyunov:2015mqj}.}

 How then do we determine if a worldsheet theory is integrable?   In the absence of general criteria for a string worldsheet theory to admit a Lax formulation, and hence classical integrability,  this is rather hard.  Disproving integrability would seem to be a more tractable problem, at least in principle. Even this task presents challenges, since it requires a  rigorous analysis of the full non-linear PDE's that arise from a string sigma model.   A more practical approach is to study particular wrapped string configurations upon which the  worldsheet theory admits a consistent one-dimensional truncation and whose equations of motion (consisting of  non-linear ODE's)  can be analysed either numerically or analytically.  Should this truncation display non-integrability then one can conclude that the parent string worldsheet theory is also non-integrable.  This strategy was applied in the papers 
\cite{Zayas:2010fs,Basu:2011dg,Basu:2011fw,Basu:2012ae,Stepanchuk:2012xi,Giataganas:2013dha,Chervonyi:2013eja,Giataganas:2014hma,Asano:2015eha,Asano:2015qwa,Ishii:2016rlk,Panigrahi:2016zny,Basu:2016zkr,Giataganas:2017guj,Roychowdhury:2017vdo,Nunez:2018ags}.  
The method is to propose a string soliton with $l$-degrees of freedom, write its classical equations of motion, find  simple solutions for $(l-1)$ of these equations and replace the solutions in a fluctuated version the last equation. One arrives to a linear second order differential equation known as the normal variation equation (NVE) which takes the form $y'' +{\cal B} y'+{\cal A}y=0$.  The  existence of closed form or Liouvillian solutions depends on the characteristics of a combination of the functions ${\cal A}$ and ${\cal B}$ and its derivatives. The work of Kovacic \cite{Kovacic85} gives criteria for a Liouvillian solution to exist and even provides a algorithmic construction of such solutions (we review this technology in Appendix \ref{sec:Tech}).  When the NVE is not Liouvillian integrable then we can conclude the string worldsheet theory is also non-integrable.

Our goal in this paper is to use both analytic and numerical techniques to illustrate that a wide class of gravity duals to ${\cal N}=2 $ SCFTs are non-integrable. These SCFTs were introduced by Gaiotto in  \cite{Gaiotto:2009we}. We will show the non-integrability by examining string winding configurations in Type IIA Gaiotto-Maldacena spacetimes \cite{Gaiotto:2009gz}. Such spacetimes are classified  by a single function $V(\sigma,\eta)$ that solves a Laplace equation with a given charge density.   With appropriate boundary conditions, which ensure regularity of the spacetime,  the quiver gauge theory can be directly extracted from the charge density.   Two choices of $V(\sigma,\eta)$ will be studied in greater detail; first is the case  of the Sfetsos-Thompson background   \cite{Sfetsos:2010uq} where $V=V_{ST}$  corresponds to the spacetime obtained as the non-Abelian T-dual of $AdS_5 \times S^5$. The second corresponds to the Maldacena-Nunez solution  $V=V_{MN}$ obtained in \cite{Maldacena:2000mw}.   Though neither of these strictly satisfy the boundary conditions, the two solutions are somewhat fundamental. {Indeed,   any such IIA Gaiotto-Maldacena spacetime can be build of an appropriate superposition of $ V_{MN}$ potentials--see for example \cite{ReidEdwards:2010qs}, \cite{Aharony:2012tz}. Also,  a particular scaling limit  of a generic Gaiotto-Maldacena solution approximates to the Sfetsos-Thompson one, see \cite{Macpherson:2015tka}}.      We will see that rotating and wrapped strings in the background generated by $V_{MN}$, fail to be integrable.  However no-evidence of non-integrability is found for rotating and  wrapped strings in the  space time generated by $V_{ST}$.   In fact it has long been known that the non-Abelian T-duality that  relates this background to   $AdS_5 \times S^5$ is, at least in the bosonic sector, a canonical transformation    \cite{Lozano:1995jx}.  Hence one would anticipate from the outset that  strings in the Sfetsos-Thompson spacetime are classically integrable.  This integrability was proven by Borsato and Wulff \cite{Borsato:2016pas,Borsato:2017qsx} who explained how such dualities act on the Lax connection of the $\mathbb{Z}_4$ graded super-coset formulation of the $AdS_5 \times S^5$ superstring.

 The structure of the rest of the paper is as follows:    In  section \ref{sec:Simple} we will place the general strategy of the paper---described above-- in the context of the simple example of strings in a background obtained by non-Abelian T-duality on $\mathbb{R} \times S^3$. Both numerical and analytic approaches used to diagnose non-integrability turn out to be negative, a result which we explain by explicitly illustrating the Lax-formulation of the dynamics.\footnote{Despite this example, it should be emphasised that failure to detect non-integrability in a particular truncation does not in general imply integrability is present.}  This study is complemented with the material in Appendix  \ref{sec:Tech}, where we review the analytical techniques required.   In section  \ref{sec:GM} we move to the study of wrapped string configurations in gravity duals to ${\cal N}=2 $ SCFTs as outlined in the strategy described  above.   We find that in generic Gaiotto-Maldacena backgrounds, dual to ${\cal N}=2$ SCFTs, the string soliton's equations of motion have non-Liouvillian solutions. As a consequence, the background and the dual SCFT are non-integrable.  We conclude in Section \ref{concl}. Other appendices complement the presentation, making our technical results more solid.

   \section{A warm up: the non-Abelian T-dual of $R\times S^3$}\label{sec:Simple} 
    In this paper we will consider bosonic strings propagating in a curved target spacetime  endowed with a metric $G_{\mu \nu}$ and an NS two-form  $B_{\mu \nu}$ described by the non-linear sigma-model\footnote{We denote the spatial worldsheet coordinate as $\tilde\sigma $  to avoid conflict later.  Light cone coordinates are $\sigma^\pm = \tau \pm \tilde{\sigma}$ and we work in units where $\alpha' =1$.   Hermitian generators of  $\frak{su}(2)$ are $T_a = \frac{\tau^a}{2}$ such that $l_\alpha = - i g^{-1} \partial_\alpha g \equiv l_\alpha^a T_a $ with   $l_\alpha^a$ real. We choose $\epsilon^{01}= -1$. } 
  \be
  \begin{aligned}
S&=\frac{1}{ 4 \pi\alpha'}\int d\tau d\tilde{\sigma}\left[G_{\mu\nu}\eta^{\alpha\beta} +B_{\mu\nu}\epsilon^{\alpha\beta}   \right] \partial_\alpha X^\mu \partial_\beta X^\nu \label{xxz1}  \\ 
 &=\frac{1}{ \pi\alpha'}\int d\tau d\tilde{\sigma} \left[G_{\mu\nu}  +B_{\mu\nu}   \right] \partial_+ X^\mu \partial_- X^\nu \  ,
 \end{aligned}
\ee
supplemented by the Virasoro constraints that require the vanishing of  the stress tensor $T_{\alpha \beta}$:
\be
G_{\mu \nu}  \dot{X}^\mu X^{\prime \nu} \approx0  \ , \quad  \quad G_{\mu \nu} ( \dot{X}^\mu \dot{X}^{ \nu} +  X^{\prime \mu } X^{\prime \nu }  )  \approx0 \ . 
\ee

 One of the simplest such theories is given by a target space $\mathbb{R}\times S^3$ with no NS flux.   Introducing an $SU(2)$ valued worldsheet field $g(\tilde\sigma ,\tau)$ to parametrise the $S^3$, of radius $\kappa$, and $\frak{su}(2)$  current $l_\alpha = - i g^{-1} \partial_\alpha g$  the non-linear sigma model can be cast as 
\be\label{eq:act1}
S = \frac{1}{4 \pi  }  \int d\tau d\tilde{\sigma}\eta^{\alpha\beta}\left(  \frac{\kappa^2}{2} \textrm{tr}(l_\alpha l_\beta)  - \partial_\alpha X_0 \partial_\beta X_0  \right) \ . 
\ee
Here we are working classically and have fixed the world sheet metric to be flat Minkowski space.  The field $X_0$ is decoupled and free however by exploiting (fixing)  residual time-reparametrisations it can be placed in static gauge  $X_0  = E \tau$.  This leaves only undetermined the dynamics of $g$,  for which we express the equations of motion and Bianchi identity of the current   $l_\alpha$   in terms of the flatness of an  $\frak{su}(2)^\mathbb{C}$  valued Lax connection  
\be
\Lax_\pm [z] = \frac{i}{1\mp z}  l_\pm \ , \quad  [\partial_+ +  \Lax_+ , \partial_- + \Lax_-] = 0 \ , 
\ee 
leading to the integrability of the theory.   The Hamiltonian analysis here is   somewhat involved since the imposition of static gauge and the Virasoro constraints necessitate  the introduction of Dirac brackets (a particularly comprehensive review of this system can be found in \cite{Vicedo:2011zz}).   
 
The sigma-model defined by eq.~\eqref{eq:act1} has an $SU(2)_L \times SU(2)_R$ global symmetry arising from the isometries of the $S^3$.  We can T-dualise the $SU(2)_L$ symmetry using a non-Abelian extension of the Buscher procedure  \cite{delaOssa:1992vci}.  This is achieved by introducing $\frak{su}(2)$ valued gauge fields to gauge the isometry and Lagrange multipliers $v^a$ that enforce the flatness of the corresponding field strength.   After integrating out the non-propagating gauge fields one obtains a sigma-model of the form  
  \be\label{eq:act2}
S = \frac{1}{  \pi  }  \int d\tau d\tilde{\sigma}   \, \partial_+ v^a (M^{-1})_{ab} \partial_- v^b  -  \partial_+ X_0 \partial_- X_0    \ , \quad M_{ab} = (4 \kappa)^2 \delta_{ab} + 4 f_{ab}^c v_c \ . 
\ee
 After making a coordinate transformation 
  \be
v^1 =  2 \sqrt{2} \kappa^2  r \cos \chi \ , \quad v^2 =  2 \sqrt{2} \kappa^2 r  \sin \chi \cos \xi   \ , \quad v^3 = 2 \sqrt{2} \kappa^2  r \sin \chi \sin \xi    \ , 
\ee
we can express the target space geometry of the dual theory as
\be
ds^2 = -( dX_0)^2 + \frac{\kappa^2}{2} \left[  dr^2 + \frac{r^2}{1+r^2} \left( d\chi^2 + \sin^2\chi d\xi^2 \right) \right] \ , \quad B_2 = - \frac{\kappa^2}{2} \frac{r^3}{1+r^2} \sin\chi d\chi\wedge d\xi \ . 
\label{rxs3natd}
\ee  
There is also a non-constant dilaton coming from the Gaussian elimination of gauge fields but this will play no role in the classical analysis we continue with here.

From this construction it is natural to expect that the T-dual theory eq.~\eqref{eq:act2} is integrable and this is indeed the case as we shall explain shortly.  Let us first examine how the diagnostic tools developed in references  \cite{Zayas:2010fs} -\cite{Roychowdhury:2017vdo} and further explained in Appendix \ref{sec:Tech} can be applied and will fail to detect non-integrability. 

We consider a wrapped string configuration specified by the ansatz 
\be
X_0 = E  \tau  \ , \quad r = r(\tau) \ , \quad  \chi = \chi(\tau) \ , \quad \xi = k \tilde\sigma  \ . 
\ee
The equations of motion for this reduced system are 
\be\label{eq:eqmred}
\begin{aligned}
\ddot{r} &= \frac{r \left(-k^2 \sin ^2 \chi  +k r \left(r^2+3\right) \dot\chi   \sin  \chi  +\dot\chi{}^2\right)}{\left(r^2+1\right)^2} \\ 
\ddot{\chi} &= -\frac{k r \sin  \chi  \left(k \left(r^2+1\right) \cos  \chi +\left(r^2+3\right) \dot{r} \right)+2 \dot{r} \dot{\chi} }{r^3+r}
\end{aligned}
\ee
and follow from the   Hamiltonian 
\be
\label{eq:Htrunc}
H = \frac{2 \pi  }{\kappa^2} \left(p_r^2 + \left( 1 + \frac{1}{r^2} \right) p_\chi^2  \right) + k r \sin \chi \left( \frac{ k r \kappa^2}{8 \pi }  - p_\chi\right).
\ee
The  non-trivial Virasoro constraint is 
\be
4 \pi   H - E^2 \approx 0 \ . 
\ee 
There are no secondary constraints since the derivative of the Virasoro constraint vanishes on the equations of motion.  

Let us now trial the diagnostic tests of non-integrability on this reduced system.  

The first test is numerical; we calculate the Lyapunov exponent.  Given two initial points $ X_0 $ and $ X_0+\Delta X_0 $, arbitrarily close in the phase space, the Lyapunov exponent is defined as 
\begin{eqnarray}
\hat{\lambda }= \lim_{t \rightarrow \infty} \lim_{ \Delta X_0 \to 0 } \hat{\lambda }(t) \ , \quad  \hat{\lambda }(t)  = \frac{1}{t}\log \frac{|\Delta X (X_{0},t)|}{|\Delta X (X_{0},0)|} \ .
\end{eqnarray}
This provides a quantitative measure on the rate of increase (or decrease) in the separation between two infinitesimally close trajectories in the phase space. In our analysis, for obvious reasons, we will be concerned with the largest positive Lyapunov  exponent and consider $\hat{\lambda}(t)$  measured at sufficiently late times. The largest Lyapunov  exponent  corresponding to dynamical systems exhibiting integrable trajectories in the phase space is identically zero. On the other hand, for systems exhibiting chaos $ \hat{\lambda} $ is non zero and saturates to a positive value that guarantees an exponential growth in the separation between two nearby trajectories at sufficiently late times.    We calculate numerically  $\hat{\lambda}(t)$ by evolving a set of   initial conditions that identically set the Hamiltonian constraint equal to zero, $ H=0 $.   The result is displayed  as the $\epsilon=0$ plot in  Figure  \ref{Lyapunov}. This indicates a vanishing Lyapunov exponent thus no evidence of non-integrability.

To use the analytic tests of non-integrability \cite{Basu:2011fw}, we consider the normal variational equation (NVE)  to a seed solution to the equations  eq.~\eqref{eq:eqmred}:
\be
r(\tau) = a \tau + b \ , \quad \chi(\tau) =0  \ .  \label{vvxx}
  \ee
  We then fluctuate $\delta \chi =0+  \epsilon f(\tau)$ and take the equations to leading order in $\epsilon$: 
 \be
0=  k \left(\frac{2 a}{(a \tau +b)^2+1}+a+k\right) f(\tau) +\frac{2 a}{(a \tau +b)^3+a \tau +b} \dot{f}(\tau) + \ddot{f} (\tau )  \ . 
 \label{estaxx}\ee 
Defining $    a \tau +  b  =  \tilde\tau $,   $a \hat{k} =  k $  and $f(\tau) = \frac{\sqrt{1+ \tilde{\tau}^2}}{\tilde{\tau}} \psi(\tilde{\tau}) $ the NVE can be placed in normal form  
 \be
  \psi''(\tilde{\tau})  + V(\tilde{\tau})  \psi= 0 \ , \quad  V(\tilde{\tau}) =  \frac{3   }{\left(\tilde{\tau}^2+1\right)^2}+\frac{2  \hat{k} }{\tilde{\tau}^2+1}+ \hat{k}  (1+ \hat{k} ) \ .
\label{estaxxx} \ee 
This last equation, can be readily integrated. In fact, one can check that the function $V(\tilde{\tau})$ satisfies the necessary conditions to be Liouville integrable discussed in Appendix \ref{sec:Tech}. See in particular the analysis in Appendix \ref{manaza}.
Indeed, we can find two  independent explicit solutions of the NVE, 
 \be
 f_1(\tau) =\frac{ \hat{k} \cos \left(\tilde \tau  \sqrt{x}\right)}{\tilde \tau }-\sqrt{x} \sin \left(\tilde \tau  \sqrt{x}\right)  \ , \quad  f_2(\tau) = \frac{\hat{k} \sin \left(\tilde \tau  \sqrt{x}\right)}{\tilde \tau }+\sqrt{x} \cos \left(\tilde \tau  \sqrt{x}\right)
\label{liouvillianNATD} \ee
where  $x=  \hat{k}  (1+ \hat{k} )$.   Whilst this solution may not be completely trivial it is composed of   appropriate elementary building blocks  and is Liouvillian.  
 This analytic test  also provides no evidence of non-integrability.   Actually, as we discuss now, the integrability of the system can be proven.

Indeed, the reason both tests for non-integrability returned negative results is that the Hamiltonian in eq.~$\ref{eq:Htrunc}$ is, in fact, integrable in the Liouvillian sense.  In addition to the Hamiltonian in eq.~\eqref{eq:Htrunc}, one can readily check that a second conserved charge is given by 
\be
Q = p_r^2  + \frac{p_\chi^2}{r^2} + \frac{ k \kappa^2}{2 \pi} \left( p_r \cos  \chi  - \frac{ p_\chi}{r} \sin \chi \right).
\ee

To make the integrability of eq.~\eqref{eq:Htrunc} we could here proceed to give a Lax pair of matrices $\{ \mathbb{L}, \mathbb{M} \}$ obeying $d_\tau \mathbb{L} = [ \mathbb{L}, \mathbb{M} ] $ from which the conserved charges of $H$ and $Q$ are obtained via $Tr(\mathbb{L}^n)$.  The explicit form of the Lax matrices we found is not very enlightening and so we don't present it here. Instead, it is more powerful to see that parent two-dimensional theory defined by the T-dual action in eq.~\eqref{eq:act2} is itself integrable.   The equations of motion for the fields $v^a$ can be packaged into a Lax form   
\be
\begin{aligned} 
\widehat{\Lax}_\pm [z] = \frac{1}{1\mp z}  \widehat{l}_\pm \ , \quad  [\partial_+ +  \widehat{\Lax}_+ , \partial_- +\widehat{\Lax}_-] = 0 \ , \\  
 \widehat{l}_+ =  -4 (M^{-T}\partial_+ v)^a T_a \ ,  \quad  \widehat{l}_- =   4  (M^{-1}\partial_- v)^aT_a \ . 
\end{aligned} 
 \ee
 This Lax formulation \cite{Borsato:2016pas} follows from the mapping $l_\pm  \to  \widehat{l}_\pm$ between world-sheet derivatives obtained as a consequence of the Buscher procedure and  which defines a canonical transformation map between the starting theory  and its dual \cite{Lozano:1995jx}.  Of course one should be concerned about the transformation of the Virasoro constraints between the two theories, however it is easy to see using the definition of the matrix $M_{ab}$ in eq.~\eqref{eq:act2} that, 
 \be
 \frac{1}{2} \partial_{\pm}v^T  ( M^{-1} + M^{-T} )\partial_{\pm}v  = \kappa^2  \widehat{l}_\pm \cdot \widehat{l}_\pm  \ , 
 \ee  
   and hence the Virasoro constraints are equivalent to
   \be
    \kappa^2  \widehat{l}_\pm \cdot \widehat{l}_\pm  \approx  (\partial_\pm X_0)^2 \ . 
   \ee
These follow from the Virasoro constraints of the starting theory by the canonical transformation  $l_\pm  \to \widehat{l}_\pm$.

 Below, we use the same diagnostic tool kit to study non-integrability in a set of holographic duals to ${\cal N}=2$ SCFTs.  These Type IIA backgrounds share some of the structure of the background described around eq.(\ref{rxs3natd}).   The non-Abelian T-dual of $AdS_5 \times S^5$ however stands out as a unique example of these geometries since it provides an integrable theory \cite{Borsato:2016pas,Borsato:2017qsx}  (see appendix \ref{sec:S5int}  for an explicit construction).\footnote{For the non-Abelian T-dual of the principal chiral model a description of the Lax was provided in \cite{Sfetsos:2013wia} and the Lax pair provided by \cite{Borsato:2016pas,Borsato:2017qsx} follows, in a rather circuitous route, from one introduced in \cite{Kawaguchi:2014qwa}.}   In contrast we will see that generic backgrounds dual to ${\cal N}=2$ SCFTs will  exhibit non-integrability.
 
     \section{Integrability of  ${\cal N}=2$ conformal field theories}
    \label{sec:GM}
     
In this section, we study the (non)-integrability of  a  generic family of ${\cal N}=2$ SCFTs. Our results are presented in the language of
the  holographic string dual to these  SCFTs. 

The strategy  we adopt follows the ideas in \cite{Basu:2011di},  \cite{Basu:2011fw}: we propose a string configuration, 
and study its classical equations of motion following from the action of eq.(\ref{xxz1}). 
We shall demonstrate  analytically the non-integrability of these equations (non-existence of Liouvillian solutions as described in Appendix \ref{sec:Tech}) and show the presence of chaos in the dynamical evolution. The existence of one such string configurations rules out the integrability of the whole CFT.

The super conformal field theories we focus our attention on, were introduced by Gaiotto in \cite{Gaiotto:2009we}.
We start by summarising the holographic string dual to these conformal theories.
The general form of Type IIA backgrounds was first presented by Gaiotto and Maldacena in 
\cite{Gaiotto:2009gz}. These  solutions  
are completely determined  in terms of a potential function $V(\sigma,\eta)$. 
Denoting
\begin{equation}
\dot{V}=\sigma \partial_\sigma V, \;\;\;\ddot{V}= \sigma^2\partial^2_\sigma V+\sigma \partial_\sigma V;\;\;\; V'=\partial_\eta V, \;\;\; V''=\partial^2_\eta V,\nonumber
\end{equation}
one can write the Type IIA 
background  as
\begin{eqnarray}
& & ds_{IIA,st}^2=\alpha'(\frac{2\dot{V} -\ddot {V}}{V''})^{1/2}
\Big[  4 AdS_5 +\mu^2\frac{2V'' \dot{V}}{\Delta} 
{d \Omega^{2}_2(\chi,\xi)}+\mu^2\frac{2V''}{\dot{V}}  
(d\sigma^2+d\eta^2)+ \mu^2\frac{4V'' \sigma^2}{2\dot{V}-\ddot{V}} 
d{\beta}^2 \Big], \nonumber\\
& & A_1=2\mu^4\sqrt{\alpha'}
\frac{2 \dot{V} \dot{V'}}{2\dot{V}-\ddot{V}}d{\beta},\;\;\;\; 
e^{4\phi}= 4\frac{(2\dot{V}-\ddot{V})^3}{\mu^{4}V'' \dot{V}^2 \Delta^2}, 
\quad {\Delta = (2 \dot{V} - \ddot{V}) V'' + (\dot{V}')^2} \ ,  \nonumber \\
& & B_2=2\mu^2\alpha' (\frac{\dot{V} \dot{V'}}{\Delta} -\eta) 
d\Omega_2,\;\;\; {C}_3={-} 4\mu^4 \alpha'^{3/2}
\frac{\dot{V}^2 V''}{\Delta}d{\beta} \wedge d\Omega_2.
\label{metrica}
\end{eqnarray}
The radius of the space is $\mu^2\alpha'=L^2$. We use 
the  two-sphere metric $d \Omega^{2}_2(\chi,\xi)=d\chi^2+\sin^2\chi d\xi^2$, with corresponding volume 
form $d\Omega_{2}= \sin\chi d\chi \wedge d\xi$. The usual definition
$F_4= dC_3 + A_1\wedge H_3$ is also used.

To write backgrounds in this family, one should find the function $V(\sigma,\eta)$ that solves
a Laplace problem with a given charge density $\tilde{\lambda}(\eta)$ and boundary conditions,
\begin{eqnarray}
& & \partial_\sigma[\sigma \partial_\sigma V]+\sigma \partial^2_\eta V=0,\nonumber\\
& &\tilde{ \lambda}(\eta)= \sigma\partial_\sigma V(\sigma,\eta)|_{\sigma=0}. \;\;\;\; \tilde{\lambda}(\eta=0)=0,\;\; \tilde{\lambda}(\eta=N_c)=0.
\label{ecuagm1}
\end{eqnarray}
The boundary condition at $\eta = N_c$ ensures that the corresponding SCFT quiver will have finite length, in this work we will also want to relax this boundary condition such that e.g. the Maldacena-Nunez solution is incorporated.

For the purposes of the classical analysis of our bosonic sigma model string solution, we need only
the metric and the $B_2$-field in the configuration of eq.(\ref{metrica}). We introduce the notation
\begin{eqnarray}
& & ds^2=4f_1(\sigma,\eta)  AdS_5 + f_2(\sigma,\eta)(d\sigma^2+d\eta^2)+ f_3(\sigma,\eta) d\Omega_2(\chi,\xi)+ f_4(\sigma,\eta) d\beta^2,\nonumber\\
& & B_2= f_5(\sigma,\eta) \sin\chi d\chi \wedge d\xi.
\label{mmmaaa}
\end{eqnarray}
In what follows, we set $\alpha'=L=\mu=1$. Comparing with eq.(\ref{metrica}), 
the functions $f_i(\sigma,\eta)$ are,
\begin{eqnarray}
& & f_1=(\frac{2\dot{V} -\ddot {V}}{V''})^{1/2},\;
f_2= f_1 \frac{2V''}{\dot{V}} ,\; f_3=f_1 \frac{2V'' \dot{V}}{\Delta} ,\; 
f_4=f_1 \frac{4V'' \sigma^2}{2\dot{V}-\ddot{V}} ,\; f_5=2(\frac{\dot{V} \dot{V'}}{\Delta} -\eta) . 
\label{functkk}
\end{eqnarray}
Let us now propose a string configurations and study its dynamical evolution.

\subsection{Study of strings in Gaiotto-Maldacena backgrounds}
We consider
a string that sits in the center of $AdS_5$ rotates and wraps on the following coordinates ($\tau,\tilde{\sigma}$ are the world-sheet coordinates),
\begin{eqnarray}
& & t=t(\tau),\;\;\; \sigma=\sigma(\tau),\;\; \eta=\eta(\tau),\;\;\;\chi=\chi(\tau);\;\;\;\; \xi=k\tilde{\sigma},\;\;\;\; \beta=\lambda \tilde{\sigma}.\label{confix}
\end{eqnarray}
With $(k,\lambda)$ being integer numbers that indicate how many times the string wraps each of the corresponding directions.

To study the equations of motion, we write an effective Lagrangian using eq.(\ref{xxz1}) and the associated Virasoro constraint.
For the  configuration in eq.(\ref{confix}), we have,\footnote{In the following, we denote with a  dot the $\tau$-derivative, which should not be confused with the 
$\sigma\partial_\sigma$ derivative defined above.}
\begin{eqnarray}
& & L=4 f_1 \dot{t}^2-f_2(\dot{\sigma}^2+\dot{\eta}^2)+ f_3(k^2\sin^2\chi -\dot{\chi}^2) +f_4 \lambda^2 +2 k f_5 \dot{\chi} \sin\chi. \label{virasorolag}\\
& & T_{\tau\tau}=T_{\tilde{\sigma}\tilde{\sigma}}=-4 f_1 \dot{t}^2 + f_2(\dot{\sigma}^2+\dot{\eta}^2)+ f_3(k^2\sin^2\chi +\dot{\chi}^2) +f_4 \lambda^2 =0.\;\;\; T_{\tilde{\sigma}\tau}=0.\nonumber
\end{eqnarray}
There is also an  effective Hamiltonian that reads,
\be
- H= -\frac{p_t^2}{16 f_1 }+ \frac{p_\sigma^2+p_\eta^2}{4 f_2} 
+\frac{1}{f_3}
(\frac{p_\chi}{2} + k f_5 \sin\chi)^2 + k^2 f_3 \sin^2\chi +\lambda^2 f_4.\label{hamiltonianxx}
\ee
The equations of motion derived from the effective Lagrangian are,
\begin{eqnarray}
& &  f_1 \dot{t}=E.\label{teq}\\
& & f_3 \ddot{\chi}= -f_3 k^2 \cos\chi\sin\chi +k \sin\chi (\dot{\eta}\partial_\eta f_5 +\dot{\sigma} \partial_\sigma f_5)-\dot{\chi} (  
\dot{\eta}\partial_\eta f_3 +\dot{\sigma} \partial_\sigma f_3 ).\label{eqchiz}\nonumber\\
& & f_2\ddot{\sigma}\!=\!-\dot{\eta}\dot{\sigma}\partial_\eta f_2 -\!2 \frac{E^2}{f_1}\partial_\sigma \log f_1 +\!\frac{1}{2}(\dot{\eta}^2-\dot{\sigma}^2)\partial_\sigma f_2+ \!\!
\frac{1}{2}(-k^2\sin^2\chi +\dot{\chi}^2)\partial_\sigma f_3 -\frac{\lambda^2}{2} \partial_\sigma f_4 \!-\!k \sin\chi \dot{\chi}\partial_\sigma f_5.\label{eqsigma}\nonumber\\
& & f_2\ddot{\eta}=\!-\dot{\eta}\dot{\sigma}\partial_\sigma f_2 -\!2 \frac{E^2}{f_1}\partial_\eta \log f_1 +\!\frac{1}{2}(-\dot{\eta}^2+\dot{\sigma}^2)\partial_\eta f_2+ \!\!
\frac{1}{2}(-k^2\sin^2\chi +\dot{\chi}^2)\partial_\eta f_3 -\frac{\lambda^2}{2} \partial_\eta f_4 \!-\!k \sin\chi \dot{\chi}\partial_\eta f_5.\label{eqeta}\nonumber
\end{eqnarray}
Here $E$ is a constant of motion. Notice that the $t-$equation of motion---the first of eqs.(\ref{teq}) was used in the other equations.

The reader can check  that the derivative of the Virasoro constraint vanishes when evaluated on the second order equations (\ref{teq}). Hence the 
constraint is a constant 'on shell'. We choose the integration constant $E$ such that $T_{\alpha\beta}=0$.

The equations (\ref{teq}) define the $\tau$-evolution of  our string configuration.  Below, 
we discuss the possibility of finding simple solutions. We also study the non-integrability of these simple solutions
and the chaotic dynamics of  the configuration proposed in eq.(\ref{confix}).
\subsection{Finding  simple solutions}
Since the equations (\ref{teq}) depend on the functions $f_1,..., f_5$ and these in turn depend on the function $V(\sigma,\eta)$, our analysis
should start by specifying the potential function $V(\sigma,\eta)$.

Inspired by the paper \cite{Itsios:2017nou}, we choose
a potential function such that expanded close to $\sigma=0$ reads,
\be
V(\sigma,\eta)= F(\eta)+ a\eta \log\sigma+\sum_{k=1}^{\infty}h_k(\eta) \sigma^{2k}.
\label{vvvccc}
\ee
This potential satisfies the Laplace equation (\ref{ecuagm1})  if,
\be
4h_1(\eta)=-F''(\eta), \;\;\;\; h_k(\eta)=-\frac{1}{4k^2}h_{k-1}''(\eta), \;\;\; k=2,3,....\label{recurrxx}
\ee
Which determines all the functions $h_k(\eta)$ in terms of $F(\eta)$. 
This potential function gives a charge density $\tilde{\lambda}(\eta)=a \eta$, as defined by eq.(\ref{ecuagm1}). Hence, it does not satisfy the second boundary condition in eq.(\ref{ecuagm1}).

Let us now  find a simple solution to eqs.(\ref{teq}).  We observe that the potential in eq.(\ref{vvvccc}) implies  that close to $\sigma(\tau)=0$, the functions
$\partial_\sigma f_i|_{\sigma=0}=0$. This suggest  a simple configuration,
\be
\sigma(\tau)=\dot{\sigma}(\tau)=\ddot{\sigma}(\tau)=0,\nonumber
\ee
that solves automatically the equation for the $\sigma$-variable, the third in eqs.(\ref{teq}).

Expanding close to $\sigma(\tau)=0$, the functions $f_i(\sigma=0,\eta)$ are  only functions of $\eta(\tau)$. The remaining two equations for the variables $\chi(\tau)$
 and $\eta(\tau)$ read,
 \begin{eqnarray}
 & & f_3 \ddot{\chi}=-f_3  k^2 \cos\chi\sin\chi +k \sin\chi \dot{\eta}\partial_\eta f_5 - \dot{\chi}   
\dot{\eta}\partial_\eta f_3 \label{eqchixx}\\
& & 
f_2\ddot{\eta}=-2 \frac{E^2}{f_1}\partial_\eta \log f_1 -\frac{1}{2}\dot{\eta}^2\partial_\eta f_2+
\frac{1}{2}(-k^2\sin^2\chi +\dot{\chi}^2)\partial_\eta f_3 \!-\!\frac{\lambda^2}{2} \partial_\eta f_4 \!-\!k\! \sin\chi \dot{\chi}\partial_\eta f_5.\label{eqetaxx}
\end{eqnarray}
With  these two equations, we follow the procedure described in \cite{Basu:2011fw}. First, consider the situation in which $\chi(\tau)=\dot{\chi}(\tau)=\ddot{\chi}(\tau)=0$. This solves eq.(\ref{eqchixx}) and using eq.(\ref{eqetaxx})
gives,
\be
f_2\ddot{\eta}=-2 \frac{E^2}{f_1}\partial_\eta \log f_1 -\frac{1}{2}\dot{\eta}^2\partial_\eta f_2
-\!\frac{\lambda^2}{2} \partial_\eta f_4 .\label{eqetanve}
\ee
With the function $V(\sigma,\eta)$ proposed  in eq.(\ref{vvvccc}), 
we find expressions for the $f_i(0,\eta)$ and $\partial_\eta f_i(0,\eta)$. 
Using these expressions the $\eta$-equation (\ref{eqetanve}) reads (we are always expanding close to $\sigma(\tau)=0$),
\be
\ddot{\eta}=(\dot{\eta}^2-E^2)
\Big( \frac{ F''-  \eta F'''}{4\eta F''}   \Big).\label{nveeta}
\ee
This equation can be easily solved for all functions $F(\eta)$. The solution is  $\eta_s(t)= E\tau + \beta$ with $(E,\beta)$ constants, compare this with eq.(\ref{vvxx}).
We use this simple solution in what follows.

Fluctuating eq.(\ref{eqetaxx}) by $\chi(t)=0+ z(t)$, 
to first order in the fluctuation we get,
\begin{eqnarray}
& & \ddot{z}(\tau) +{\cal B}\dot{z}(\tau) 
 +{\cal A} z(\tau)=0, \label{eqdiffxx}\\
& & {\cal A}=(k^2- k \dot{\eta}\frac{\partial_\eta f_5}{f_3})|_{\eta=\eta_s}, 
\;\;\;\;{\cal B}=(\dot{\eta}\partial_\eta \log f_3)|_{\eta=\eta_s}.\nonumber
\end{eqnarray}
The integrability depends on the behaviour of this last equation 
respect to Kovacic's criterium, see Appendix \ref{sec:Tech} for an explanation and application to some examples.
In terms of the function $F(\eta)$ that defines 
the function $V$ in eq. (\ref{vvvccc}), the explicit expression of the coefficients ${\cal A}, {\cal B}$ is 
\begin{eqnarray}
& & {\cal A}=k^2+ 
k \dot{\eta} \frac{\sqrt{2}}{\sqrt{a \eta F''}}
\times \frac{2 a  F'' + 2 \eta( F'')^2 +  a\eta F'''}
{(a+ 2\eta F'')}|_{\eta_s},
\nonumber\\
& & {\cal B}=\dot{\eta} \frac{1}{2\eta F''(a + 2 \eta F'')}\times 
\Big(3 a F'' +2 \eta F''^2 + a \eta F''' -2\eta^2 F'' F'''    \Big)
|_{\eta_s}.\nonumber
\end{eqnarray}
To gain more understanding of the integrability (or not) of our string soliton, 
we should study the application of Kovacic's criterium to eq.(\ref{eqdiffxx}) for different backgrounds dual to 
${\cal N}=2$ SCFTs.
In what follows, we will specify different  functions $V(\sigma,\eta)$ that define various well known backgrounds. We shall observe that whilst the generic case turns out to be non-integrable, there is a particular background---the Sfetsos-Thompson solution, for which the string soliton of eq.(\ref{confix}) is integrable. 
After that, we complement this analytic study
with the numerical study of the eqs.(\ref{teq}) or equivalently, 
those derived from  the Hamiltonian in eq.(\ref{hamiltonianxx}).
\subsection{Some interesting examples}
Below, we apply Kovacic's  procedure (discussed in detail in Appendix \ref{sec:Tech} ) to eq.(\ref{eqdiffxx}). We discuss various examples in turn by specifying   particular potential functions $V(\sigma,\eta)$.
\subsubsection{The Sfetsos-Thompson background}
 Let us start with the potential describing the Sfetsos-Thompson background, obtained by application of non-Abelian T-duality on $AdS_5\times S^5$  \cite{Sfetsos:2010uq}. The field theoretical dual to this background was discussed in \cite{Lozano:2016kum}. 

For the Sfetsos-Thompson (ST) solution, the potential function reads
\be
V_{ST}= \eta\log \sigma -\eta \frac{\sigma^2}{2} +\frac{\eta^3}{3}.\label{vstxx}
\ee
In the notation of eqs.(\ref{vvvccc})-(\ref{recurrxx}), 
we have $a=1$, $F(\eta)=\frac{1}{3}\eta^3$, $h_1=-\frac{\eta}{2}$, $h_{k>2}=0$. The coefficients ${\cal A},{\cal B}$ are,
\begin{eqnarray}
& & {\cal A}_{ST}=k^2 +2 E k\frac{(4\eta^2+3)}{(4\eta^2+1)},\;\;\; {\cal B}_{ST}=\frac{2E}{\eta(4\eta^2+1)},\nonumber
\end{eqnarray}
and eq.(\ref{eqdiffxx}) becomes,
\be
\ddot{z}+ \frac{2E}{\eta(4\eta^2+1)}\dot{z} +\left(k^2 +2 E k\frac{(4\eta^2+3)}{(4\eta^2+1)}\right) z=0.
\label{natdliouville}
\ee
This equation admits Liouvillian solutions, the system is the same, up to a rescaling of the $\tau-$coordinate, to that analysed around eq.(\ref{estaxx}). 
Indeed,  as we did with eq.(\ref{estaxx}), we can transform eq.(\ref{natdliouville}) 
into a Schroedinger-like form 
defining $ z(\tau)= e^{-\frac{1}{2}\int d\tau {\cal B(\tau)}} \psi(\tau)$,
\be
\ddot{\psi}+ V\psi=0,\;\; 
V= \frac{1}{4}(4{\cal A} -{\cal B}^2 -2 {\cal B}')=
\frac{12}{(4\tau^2+1)^2} + k^2+2k +\frac{4k}{4\tau^2+1}.\label{dadaya}
\ee
In the last line, we have set the integration constant $E=1$. This last equation can
be easily solved, as we did with eq.(\ref{estaxxx}). Transforming back to the function $z(\tau)$, 
we find that eq.(\ref{natdliouville}) admits Liouvillian
solutions, written
as a combination of trigonometric  and rational functions. See also Appendix \ref{sec:Tech} for the application of Kovacic's criteria to this case.

The presence of a string soliton that is Liouville-integrable is certainly not enough to claim the integrability of the theory. Nevertheless,  this result  together with the analysis in Section \ref{sec:Simple},  reinforce the point that the Sfetsos-Thomspson 
background is dual to an integrable CFT (see   \cite{Borsato:2016pas,Borsato:2017qsx}). In other words, the non-Abelian T-duality does not spoil the integrable character of a background-QFT pair.

Let us now study the effect of deforming the Sfetsos-Thompson solution.

\subsubsection{Deforming the Sfetsos-Thompson solution}
As anticipated, we consider `deformations' away from the Sfetsos-Thompson solution. 
To parametrise these deformations, we propose a potential written in terms of a parameter $\epsilon$,
\be
V_{def}= V_{ST}+\epsilon\left( \frac{\eta^4}{12} +\frac{\sigma^4}{32}-
\frac{\sigma^2\eta^2}{4}  \right).
\label{vdef}
\ee
This potential function $V_{def}$ satisfies the Laplace equation. But, like the Sfetsos-Thompson potential, it does not satisfy the second boundary condition in eq. (\ref{ecuagm1}). Notice that this deformation is a solution for all values of the parameter $\epsilon$.
In the notation of eq.(\ref{recurrxx}), we have
\be
F(\eta)=\frac{\eta^3}{3} +\frac{\epsilon}{12}\eta^4, \;\;\; h_1=-\frac{\eta}{2} -\frac{\epsilon}{4}\eta^2, \;\; h_2=\frac{\epsilon}{32},\;
\;\; h_{k>2}=0.
\ee
It would be interesting to study the geometrical properties of the background generated using eq.(\ref{metrica}).
One can calculate the functions
\begin{eqnarray}
& & {\cal A}_{def}=k^2+\frac{2\sqrt{2}E k}{\sqrt{2+\eta\epsilon}} 
\frac{(3+4\eta^2 + 2\epsilon \eta +4\epsilon\eta^3 +\epsilon^2 \eta^4)}
{(1+4\eta^2 +2\epsilon \eta^3)}       ,\nonumber\\
& & {\cal B}_{def}=\frac{(8+ 5\epsilon \eta -4\epsilon \eta^3 -2 \epsilon^2\eta^4)}
{2\eta(2+\epsilon\eta)(1+4\eta^2 +2\epsilon \eta^3)} E.
\label{coeffdef}
\end{eqnarray}
and plug them in eq.(\ref{eqdiffxx}). Things go quickly awry. Indeed,  
even if we simplify these expressions by expanding ${\cal A}$
and ${\cal B}$ for small values of the deformation parameter $\epsilon$,  the differential 
equation is very
hard to solve exactly and we could not find  Liouvillian solutions.  A more refined analysis that we postpone to Appendix \ref{sec:Tech} indicates that (at least for small values of the deformation parameter) the solutions to the NVE are non-Liouvillian.

\begin{figure}
\begin{center}
\includegraphics[width=0.5\textwidth]{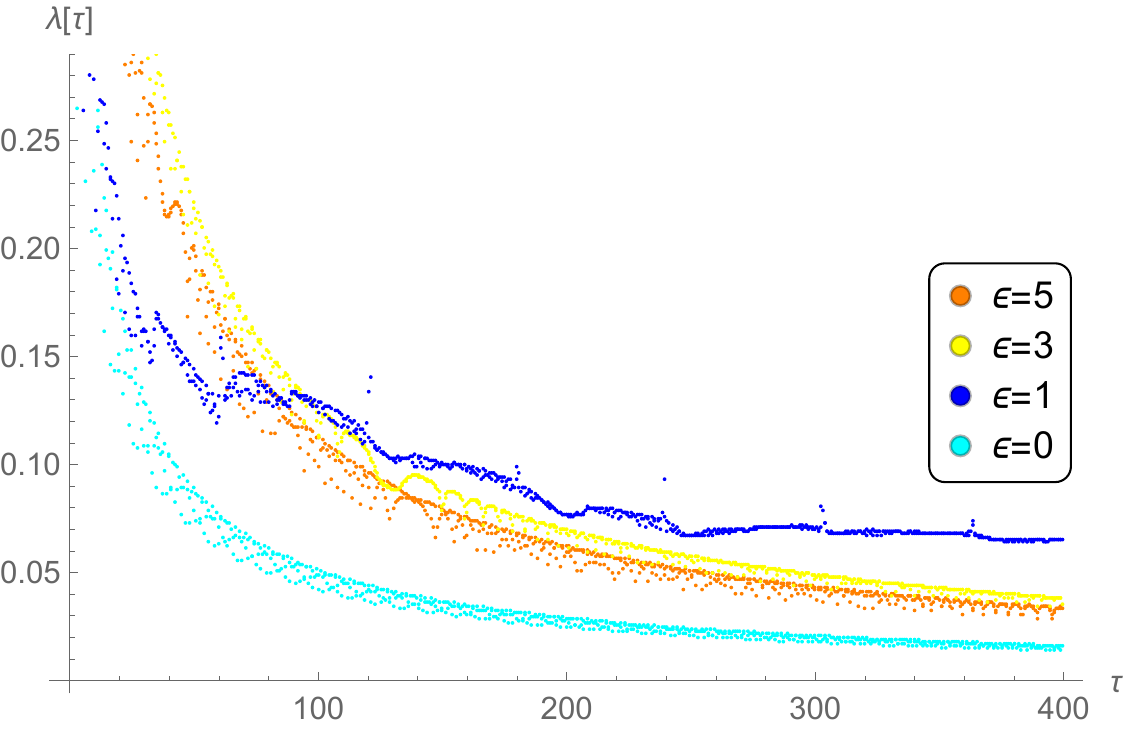}
\end{center}
\caption{For various values of the parameter $\epsilon$ defining the deformation of the Sfetsos-Thompson solution given by eq.~\eqref{vdef} we show  the  evolution of the Lyapunov coefficient (whose $t\to \infty$ limit is the  Lyapunov exponent).  The initial conditions for our analysis are, $ \chi(0)=0.5 $, $ \eta(0) =0 $, $ \dot{\chi}(0)=0.1 $, $ \dot{\eta}(0)=0.1 $ with the parameter $E$ fixed such that the Hamiltonian vanishes.  }\label{Lyapunov}
\end{figure}

We can complement this with a numerical exploration of the Lyapunov exponent $\hat\lambda$ for various values of the deformation parameter $\epsilon$.  The results of Figure~\ref{Lyapunov} indicate  that for the Sfetsos-Thompson background, the corresponding Lyapunov exhibits a sharp fall towards zero, as 
 expected from the analysis in Section \ref{sec:Simple}. On the other hand,  when $\epsilon$ is non zero  the Lyapunov exponent saturates to some positive value.
 
Together these point towards the non-integrability of the 
string soliton moving on this deformation 
of the Sfetsos-Thompson background. 
In turn this translates into the non-integrability of the associated ${\cal N}=2$ SCFT.

Below, we repeat this analysis for a solution that is qualitatively different to those in eq.(\ref{vvvccc}). 

\subsubsection{Study of the Maldacena-N\'u\~nez solution}\label{sectionmnxx}

Here, we repeat the study in the previous sections, applying it  to the solution of
\cite{Maldacena:2000mw}, for the case in which there are $N$ D4 branes in Type IIA. In this case, it is useful to work with the $\sigma$-derivative 
of the potential function $\dot{V}(\sigma,\eta)$,
\be
2 \dot{V}_{MN}(\sigma,\eta)=\sqrt{\sigma^2+(N+\eta)^2} - \sqrt{\sigma^2+(N-\eta)^2},\label{mnsolution}
\ee
and calculate all other derivatives appearing in the Gaiotto-Maldacena background,
$\ddot{V}, \dot{V'}, V''$ close to $\sigma(\tau)=0$ using an expansion of eq.(\ref{mnsolution})
\begin{eqnarray}
& & \dot{V}_{MN}(\sigma\sim 0,\eta)
\sim \frac{1}{2}\lambda(\eta) +\frac{Q(\eta)}{4}\sigma^2 +\frac{Z(\eta)}{16}\sigma^4,\nonumber\\
& & \lambda(\eta)= |N+\eta |-|N-\eta|,\;\;\; Q(\eta)=\frac{|N+\eta|}{(N+\eta)^2}-\frac{|N-\eta|}{(N-\eta)^2},\nonumber\\
& & Z(\eta)=\frac{|N-\eta|}{(N-\eta)^4} -\frac{|N+\eta|}{(N+\eta)^4}.\nonumber
\end{eqnarray}
The reader can check that $\partial_\sigma f_i|_{\sigma(\tau)=0}=0$ 
and the $\sigma$-equation in (\ref{teq})  is solved for $\sigma(\tau)=\dot{\sigma}(\tau)=\ddot{\sigma}(\tau)=0$. It can also be checked 
 that $\eta =E \tau$ solves the $\eta$-equation close to $\sigma(\tau)=0$. 
The $\chi(\tau)$-equation after a fluctuation reads as in eq.(\ref{eqdiffxx}) with,
\begin{eqnarray}
& & {\cal A}_{MN}=k^2+E k \frac{1}{(2 Q \lambda -\lambda'^2) \sqrt{2 \lambda Q}} \left[ 4 Q^2 \lambda +2Q \lambda \lambda'' -4\lambda'^2 Q  +\lambda'^2\lambda'' - 2 Q'\lambda \lambda'   \right],\nonumber\\
& & {\cal B}_{MN}= \frac{E}{2 Q \lambda (2 Q\lambda -\lambda'^2)}\left[2 Q^2 \lambda \lambda'-\lambda \lambda'^2 Q' +4 \lambda \lambda'\lambda''Q -3\lambda'^3 Q -2 Q Q'\lambda^2   \right]\label{flucmn}
\end{eqnarray}
Here again, the equation is complicated enough and we don't find Liouvillian solutions.  This strongly suggests the non-integrability of the string soliton moving on this background.  A more refined analysis, involving necessary conditions for the equations to be integrable, is given in Appendix  \ref{sec:Tech}.  There we show that, for evolution in the domain $0\leq \eta \leq N$   we can make  a redefinition that transforms $ {\cal A}_{MN}$ and $ {\cal B}_{MN}$ into  rational functions and we can see that the criteria of \cite{Kovacic85} are not satisfied.  This shows the  Liouvillian non-integrability of the NVE which in turn indicates  that the CFT is not-integrable. 


The solution of  \cite{Maldacena:2000mw} was used by the authors of  
\cite{ReidEdwards:2010qs} to construct the general solutions to the Laplace problem
in eq.(\ref{ecuagm1}).   Heuristically one can think of the general Gaiotto-Maldacena background as being provided by a superposition of multiple MN profiles.  Thus  since a single MN profile leads to non-integrability this is  a strong indication of the 
non-integrability the string soliton in a generic Gaiotto-Maldacena background. 
This translates to the non-integrability of the general Gaiotto ${\cal N}=2$ SCFT. 

%
Let us close the section by summarising the results. We  found that for  a generic background in the family of Gaiotto-Maldacena solutions, it is  easy to find a string soliton whose equations of motion lead to non-Liouvillian solutions. This implies the non-integrability of the dual ${\cal N}=2$ SCFTs. A very interesting exception is the Sfetsos-Thompson solution \cite{Sfetsos:2010uq}, obtained via non-Abelian T-duality on $AdS_5\times S^5$.  Parts of our analysis used that the potential function $V(\sigma,\eta)$ can be written as in eq.(\ref{vvvccc}). This is an important limitation to our approach that we amend with the discussion in Appendix \ref{appendixgm}.

\section{Numerical analysis}\label{section-num-an}
Here, complementing the material in previous sections,  we provide a light numerical analysis supporting the analytic study of non-integrability in some  the Gaiotto-Maldacena backgrounds.  

\subsection{The backgrounds }
We shall consider two representative  backgrounds, that capture all the features of the Gaiotto-Maldacena solutions and CFTs. We  write the associated potential functions $\dot{V}$ as a sum of $\dot{V}_{MN}$ in eq.(\ref{mnsolution}).

The first background is dual to a CFT with gauge group $SU(N)\times SU(2N)\times.... \times SU(PN)$ closed by a $SU((P+1)N)$ flavour group illustrated by the quiver:   

\tikzstyle{gauge}=[circle,draw=blue!50,fill=blue!20,thick]
\tikzstyle{flav}=[rectangle,draw=black!50,fill=black!20,thick]

\begin{center}  
 \begin{tikzpicture} \label{fig:onekinkquiver}
  \node (name1) at ( -1.25,0) [gauge] {$\,\,N\,$};
    \node (name2) at (0,0) [gauge]  {$2N$};
    \draw[-] (name1.east) -- (name2.west);
        \node (name3) at (1.5,0) [gauge]  {$3N$};
            \draw[-] (name2.east) -- (name3.west);
    
       \node (name5) at (5.5,0) [gauge] {  $P N$};
               \draw[densely dotted] (name3.east) -- (name5.west);
                           \node (name6) at (8,0) [flav] {$(P+1) N$};    
                             \draw[-] (name5.east) -- (name6.west);
    \end{tikzpicture}
\end{center}
 
 The function $\dot{V}(\sigma,\eta)=\sigma\partial_\sigma V(\sigma,\eta)$ is given by,
\begin{eqnarray}
& &  \dot{V} =  \frac{N}{2}  \sum_{n=-\infty}^{\infty} (P+1)\left[\sqrt{\sigma^2 +(\eta +P -2 n (1+P) )^2} - \sqrt{\sigma^2 +(\eta -P -2 n (1+P) )^2}   \right] +\nonumber\\
& & P\left[\sqrt{\sigma^2 +(\eta -1-P -2 n (1+P) )^2} -\sqrt{\sigma^2 +(\eta +1+P -2 n (1+P) )^2}  \right].\label{quiver1}
\end{eqnarray}

\begin{figure}
\begin{center}  
\includegraphics[width=0.3\textwidth]{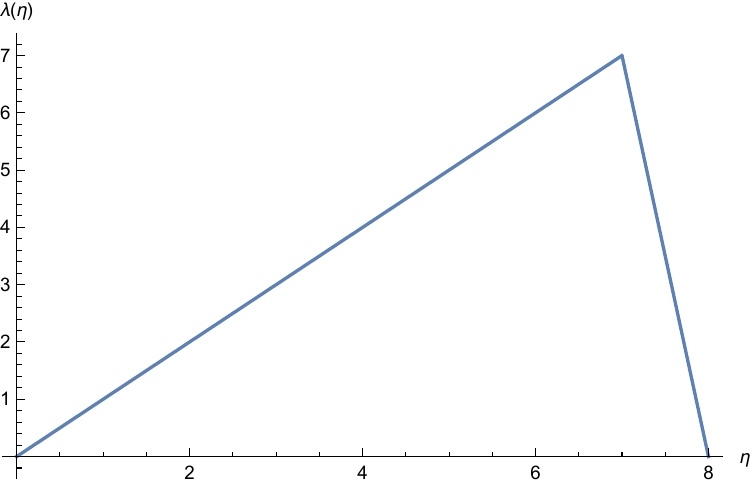} \quad \includegraphics[width=0.3\textwidth]{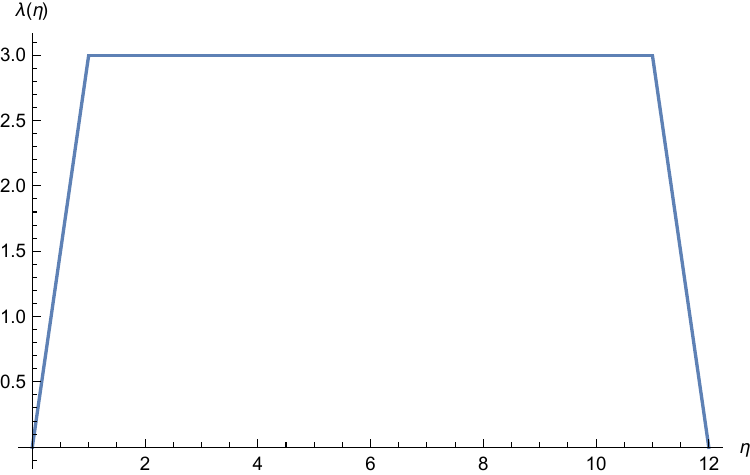} 
\caption{Charge density $\lambda(\eta)$ of a ``one-kink'' spacetime (left) and  an ``Uluru'' spacetime (right) with the values used for the numerical analysis. } 
\label{fig:chargedensity}
\end{center} 
\end{figure}

In the following we shall refer to this Gaiotto-Maldacena geometry as a ``one-kink'' spacetime due to the profile of its charge density shown on the left of    fig.~\ref{fig:chargedensity}. 
Notice that the kink where the charge density changes slope is associated to the flavour group of the quiver.  
This QFT is of particular interest since it has been proposed and studied in \cite{Lozano:2016kum}   as a {\em completion} of the  solution of \cite{Sfetsos:2010uq}.  When performing the numerical study displayed below, we implemented a cut-off on the summation over the index $n$ taking it to range $-10\leq n\leq 10$  and we took the value $P=7$.  

%
 
The second background we consider is defined by a periodic ``Uluru'' profile \cite{ReidEdwards:2010qs} with the function $\dot{V}$ given as:
\be
\dot{V}(\sigma, \eta) = \frac{N}{2} \sum_{n = -\infty}^{\infty} \sum_{l =1}^3 \sqrt{\sigma^2 + (\nu_l + n \Lambda - \eta)^2 }  -  \sqrt{\sigma^2 + (\nu_l  - n \Lambda + \eta)^2 } \ , 
\ee
 with $V'' = -\frac{1}{\sigma^2} \ddot{V}$.   Here the extra parameters   $ \Lambda = 2 K+4$, $\nu_1 = 1$, $\nu_2 = K+1$, $\nu_3 = - K -2$ define a quiver consisting of $K$ $SU(N)$ gauge group nodes terminated on each end by an $SU(N)$ flavour node:
 
\tikzstyle{gauge}=[circle,draw=blue!50,fill=blue!20,thick]
\tikzstyle{flav}=[rectangle,draw=black!50,fill=black!20,thick]
\begin{center}
 \begin{tikzpicture} 
  \node (name1) at ( -1.25,0) [flav] {$N$};
    \node (name2) at (0,0) [gauge] {$N$};
    \draw[-] (name1.east) -- (name2.west);
        \node (name3) at (1.5,0) [gauge] {$N$};
            \draw[-] (name2.east) -- (name3.west);
          \node (name4) at (4,0) [gauge] {$N$};
        \draw[densely dotted] (name3.east) -- (name4.west);
          \node (name5) at (5.5,0) [gauge] {$N$};
                    \draw[-] (name4.east) -- (name5.west);
                           \node (name6) at (7,0) [flav] {$N$};    
                             \draw[-] (name5.east) -- (name6.west);
    \end{tikzpicture}
    \end{center}
       In some sense, this is a relative of ${\cal N}=2 $ SQCD with $N_f=2N_c$.   The charge density, shown on the right of fig.~\ref{fig:chargedensity},  in this case reads
         \be
  \lambda(\eta ) = \dot{V} |_{\sigma =0} = \left\{ \begin{array}{cc} N \eta   &  0< \eta <1 \\ 
  N   &  1< \eta <K+1  \\ 
    N - N(\eta - K- 1)    &  K+1 <  \eta < K+2   \\ 
    \end{array} \right. \ . 
  \ee
 For numerical evaluations we set $N=3$ and $K=10$  and again restrict the summation  over the index $n$ ranges $-10\leq n\leq 10$. 
 
\subsection{The observables}

For both the one-kink and the Uluru spacetimes we will present three numerical plots.

\begin{figure}[h!]
\begin{center} 
  \includegraphics[width=0.25\textwidth]{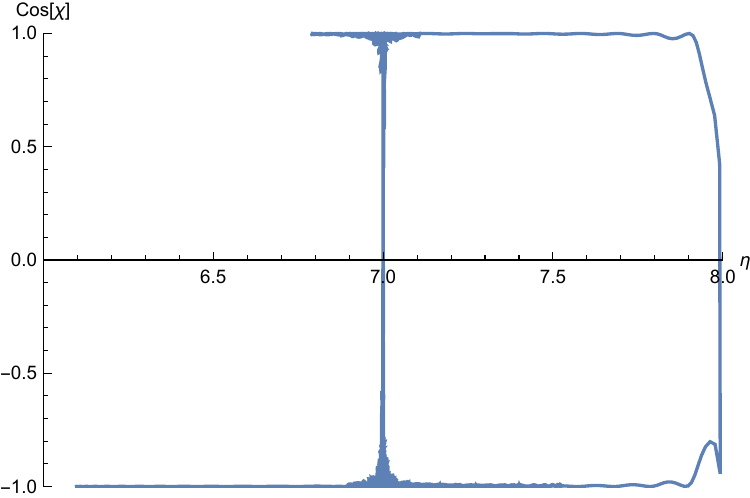}   \qquad  \qquad \qquad  \includegraphics[width=0.25\textwidth]{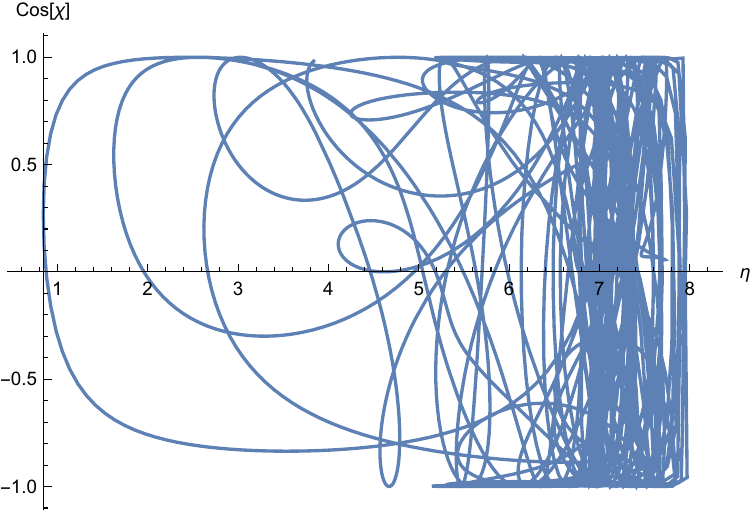}  
\end{center}
\caption{ Plots of example trajectories in the $\eta(\tau), \cos(\chi(\tau))$ plane in the one-kink space time.  Left we have $E=0.05$ and on the right $E=5.0$   } 
 \label{fig:TrajectoryOneKink}
\end{figure}

\begin{figure}[h!]
\begin{center} 
  \includegraphics[width=0.25\textwidth]{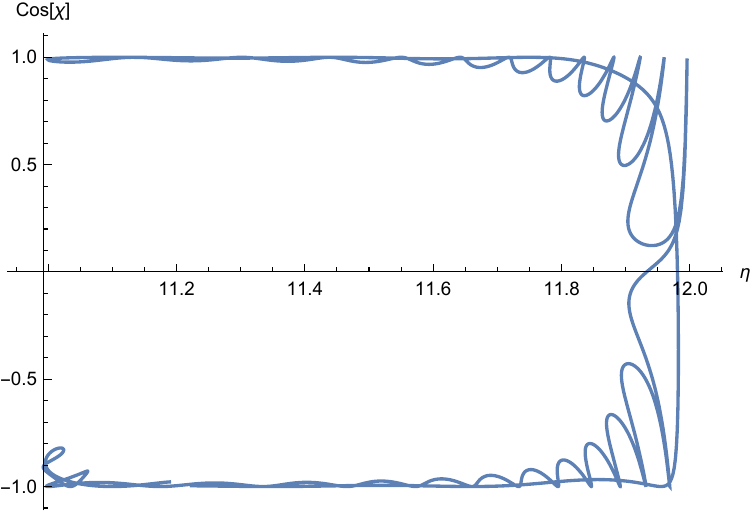}  \qquad  \qquad \qquad  \includegraphics[width=0.25\textwidth]{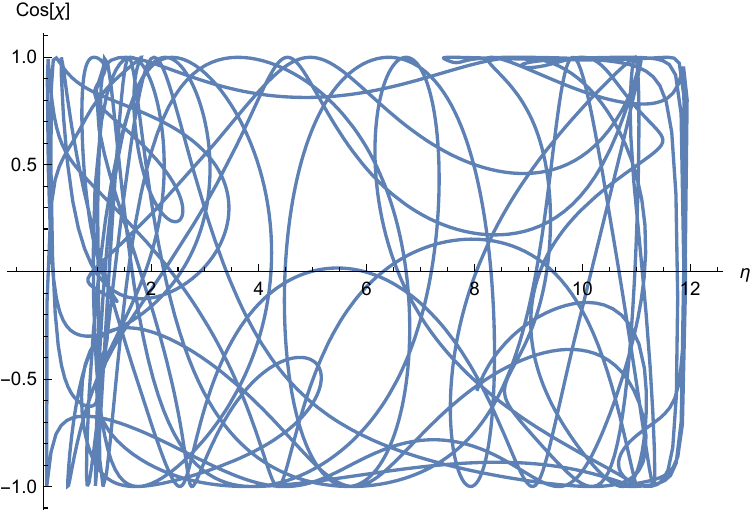}  
\end{center}
\caption{ Plots of trajectories in the $\eta(\tau), \cos(\chi(\tau))$ plane in the Uluru space time.  Left we have $E=0.1$ with see trajectory confined  to the  region of space to $11<\eta<12$  and on the right with $E=5.0$ the trajectory wildly explores all of space.    } 
 \label{fig:TrajectoryUluru}
\end{figure}

First we will consider the actual profile of trajectories in phase space.  Studying eq.(\ref{nveeta}), we found the simple solution $\eta= E\tau$. The angular coordinate was taken to be fixed at  $\chi=0$. This trajectory is certainly possible, but when reaching the end of the space (the points $\eta=8$ in the first background and the point $\eta=12$ in the second one), the trajectory should bounce back.  Here instead we will study configurations for which $\chi$ is not constant in time. We can plot numerically the trajectories in the $(\eta,\cos\chi)$-plane. We observe the trajectories are not periodic and become very disordered when raising the energy or when more generic initial conditions for $0<\chi\leq \pi$ are imposed.  The case of one-kink spacetime is shown in fig.~\ref{fig:TrajectoryOneKink} and that of the Uluru in fig.~\ref{fig:TrajectoryUluru}.

\begin{figure}[h!]
\begin{center} 
 \includegraphics[width=0.4\textwidth]{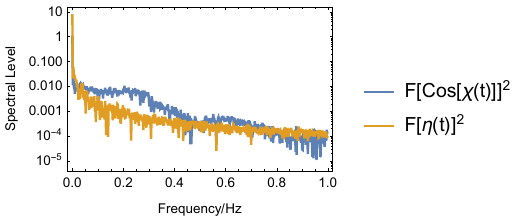}   \includegraphics[width=0.4\textwidth]{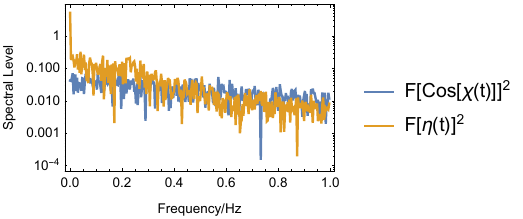} 
\end{center}
\caption{ Power spectra associated to the trajectories in the $\eta(\tau), \cos(\chi(\tau))$ plane displayed in fig.~\ref{fig:TrajectoryOneKink} for    the  one-kink space time.   Left we have $E=0.05$ and on the right $E=5.0$  } 
 \label{fig:PowerOneKink}
\end{figure}

\begin{figure}[h!]
\begin{center}  
 \includegraphics[width=0.4\textwidth]{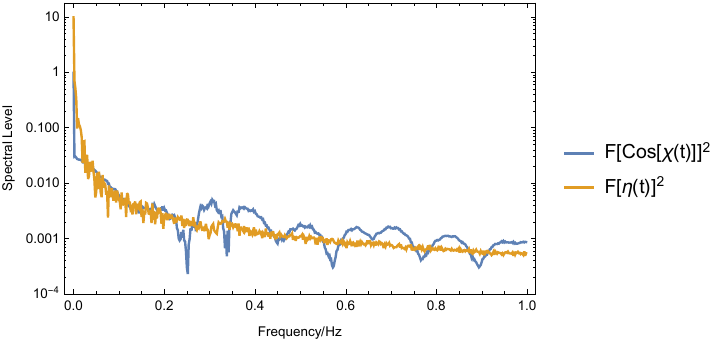} \includegraphics[width=0.4\textwidth]{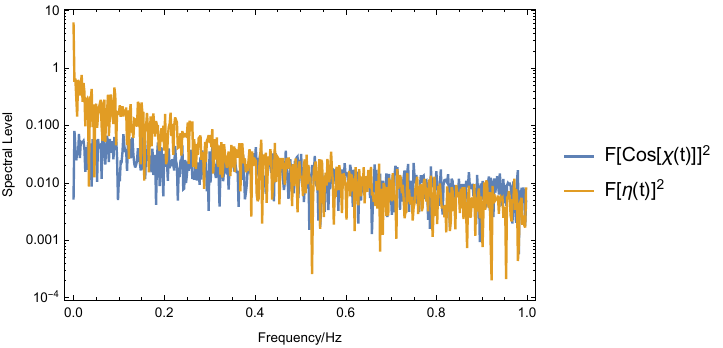}  
\end{center}
\caption{ Power spectra associated to the trajectories shown in fig.~\ref{fig:TrajectoryUluru}.   Left we have $E=0.05$ with a  see a comparatively clean spectrum and on the right $E=5.0$. } 
 \label{fig:PowerUluru}
\end{figure}

 We can then numerically perform a Fourier analysis to obtain the power spectrum \cite{Ott:2002book} of these trajectories. For periodic trajectories we should see well defined (peaked) frequencies. For non-periodic and chaotic trajectories the Fourier analysis reveals a continuum of frequencies.   This is borne out as shown in fig.~\ref{fig:PowerOneKink}  for the one-kink spacetime and   fig.~\ref{fig:PowerUluru} for the Uluru.

\begin{figure}[h!]
\begin{center} 
\includegraphics[width=0.4\textwidth]{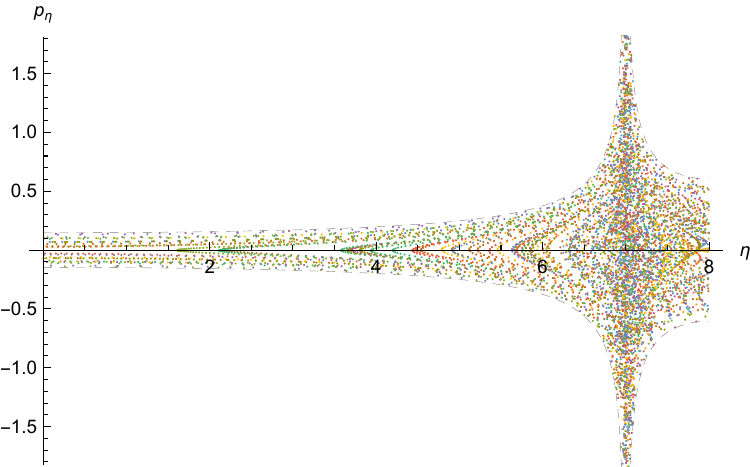}\includegraphics[width=0.4\textwidth]{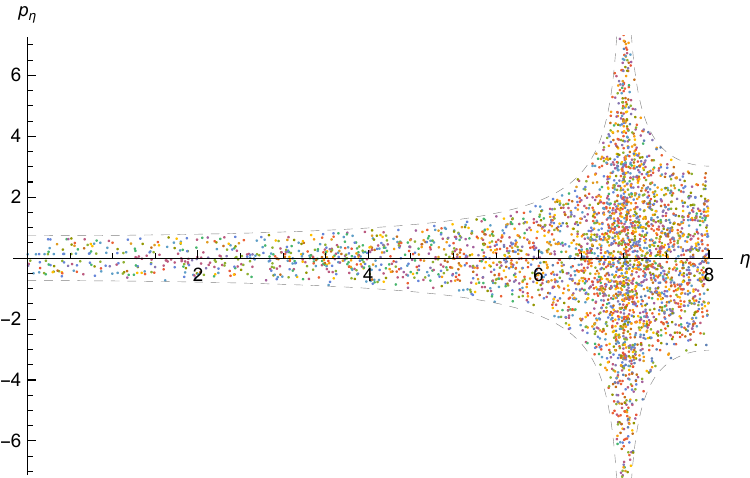}
\includegraphics[width=0.4\textwidth]{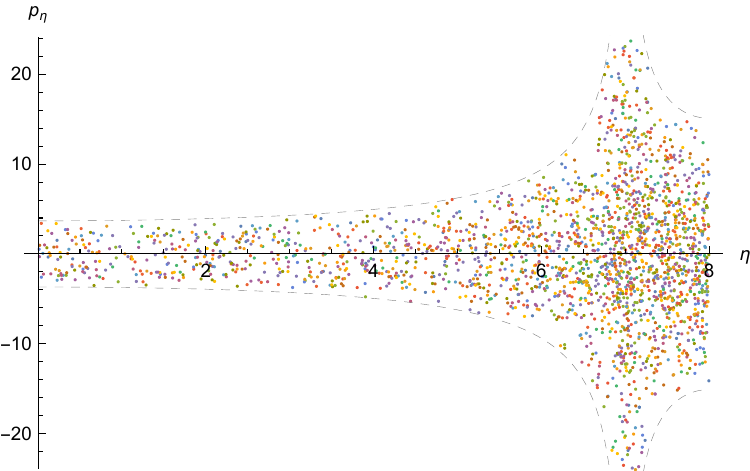}\includegraphics[width=0.4\textwidth]{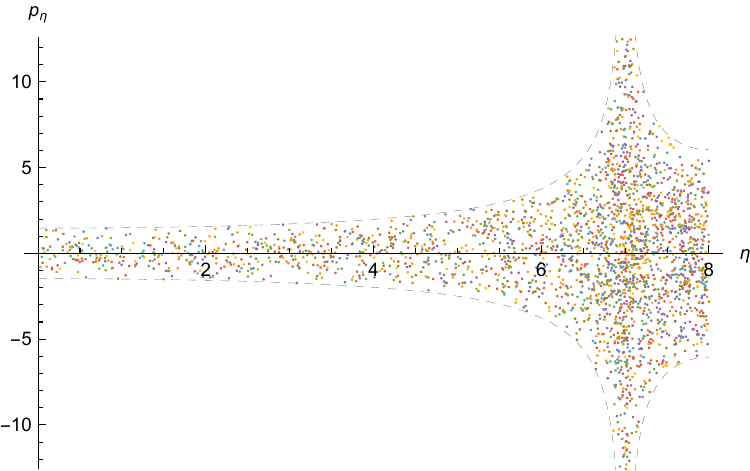}
\end{center}
\caption{  $\eta-P_\eta$ plane projections of the Poincare section at $\chi = 0$  for the one-kink spacetime.   Clockwise from top left   we vary the parameter $E = \{  0.1, 0.5, 1, 2.5\}$.  The plots fill an area bounded by  maximal value of $P_\eta$ compatible with the constraint that $H=0$ indicated with a grey dashed line.     The 100 different seed initial conditions that are numerically evolved to generate these sections are indicated by colour.  For small values of $E$ we have a perturbation around an integrable Hamiltonian (for $E=0$ the Hamiltonian is trivial and vanishing) and one sees vestiges of KAM tori which as $E$ is increased dissolve away.         } 
 \label{fig:PoincareOneKink}
\end{figure}

\begin{figure}[h!]
\begin{center} 
\includegraphics[width=0.4\textwidth]{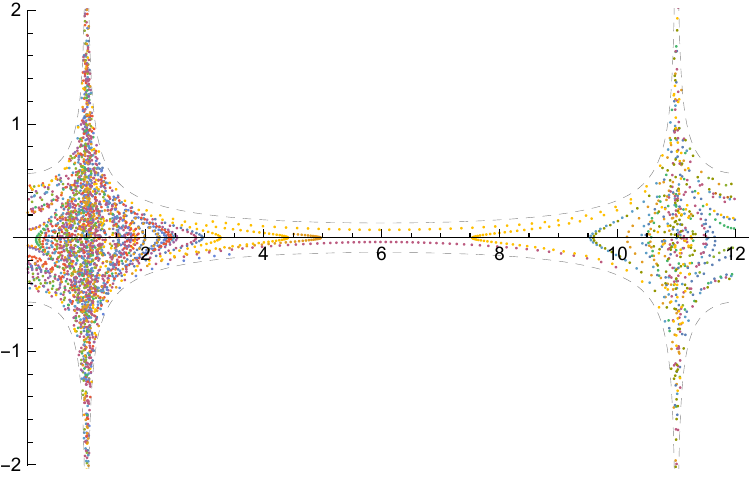}\includegraphics[width=0.4\textwidth]{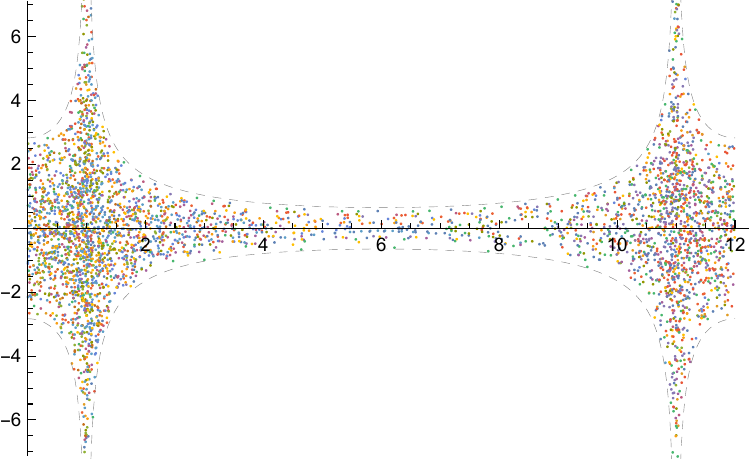}
\includegraphics[width=0.4\textwidth]{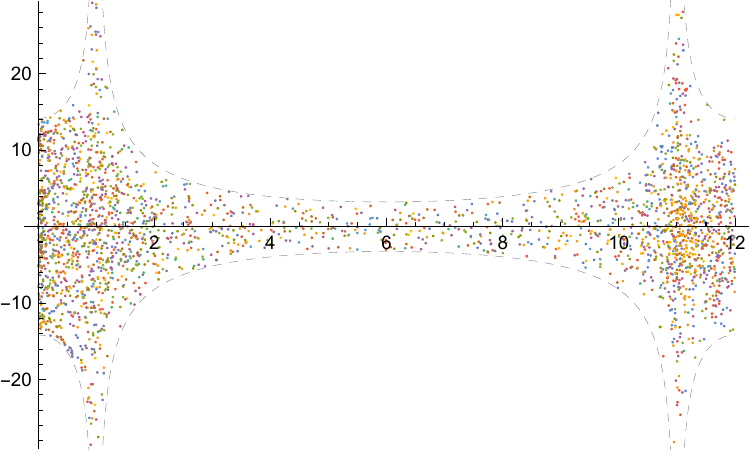}\includegraphics[width=0.4\textwidth]{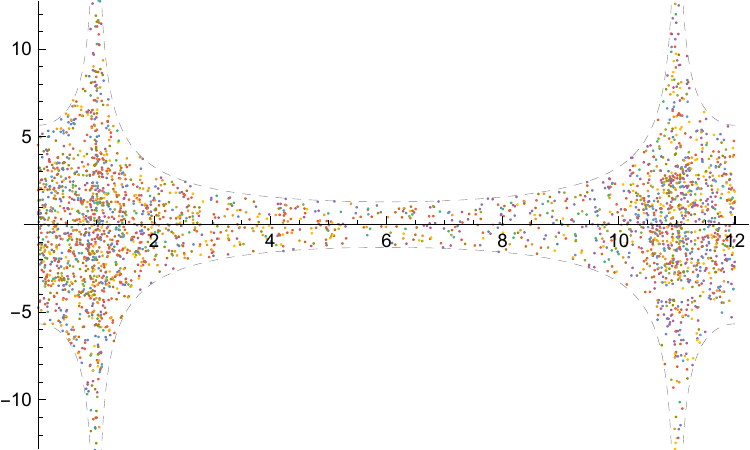}
\end{center}
\caption{   $\eta-P_\eta$ plane projections of the Poincare section at $\chi = 0$  for the Uluru spacetime.   Clockwise from top left    $E = \{  0.1, 0.5, 1, 2.5\}$.  The plots fill an area bounded by  maximal value of $P_\eta$ compatible with the constraint that $H=0$ indicated with a grey dashed line.     The 100 different seed initial conditions that are numerically evolved to generate these sections are indicated by colour.  For small values of $E$ we have a perturbation around an integrable Hamiltonian  and one sees vestiges of KAM tori which as $E$ is increased dissolve away.     } 
 \label{fig:PoincareUluru}
\end{figure}

Let us move to study the Poincar\'e sections.  Recall that an $N$-dimensional integrable system possesses $N$ independent integrals of motion that are in {involution}. That is the Poisson bracket of any two of these conserved quantities vanish.  The corresponding phase space trajectories are confined to the surface of an $N$-dimensional KAM torus (for a review see \cite{Ott:2002book}).   To learn whether a system is integrable or not, one should take cross-sections of its phase-space trajectories.  Such a cross-section is known as a Poincar\'{e} section \cite{Ott:2002book}. The KAM theorem tells us how these KAM curves will change when we perturb an integrable Hamiltonian with a deformation $ H'$. The resonant tori, for which these trajectories close on themselves, will be destroyed by this deformation,  the motion becomes seemingly random and loosing all of the  structure of the KAM curves in our Poincar\'{e} section. We fix an energy, find suitable boundary conditions and run a numerical analysis of the Poincar\'e sections, that is displayed in  fig.~\ref{fig:PoincareOneKink} for the one-kink spacetime and fig.~\ref{fig:PoincareUluru} for the Uluru.

\newpage
\begin{center} 
 {\bf{Numerical Plots for the one-kink spacetime} }
 \end{center}

\newpage

\begin{center} 
 {\bf{Numerical Plots for the Uluru spacetime} }
 \end{center}

 \newpage

\section{Conclusions}\label{concl}
Let us start with a summary of the paper. We studied the {\it non-integrability} properties of generic ${\cal N}=2$ SCFTs. The procedure we employed is based on the use of a particular classical string soliton, that rotates and wraps around various compact dimensions associated with the R-symmetry of the CFT. The corresponding field theory operator should have large dimension and R-charges. The study of the Hamiltonian system describing the constrained dynamics of the string, reduce to two nonlinear and coupled ordinary differential equations. We have discussed the presence or not of Liouvillian solutions for these equations using well-established mathematical techniques. We found that non-integrability is generic among the very large family of ${\cal N}=2$ SCFTs. Nevertheless, there is one notable exception
that is the field theory defined holographically by the Sfetsos-Thompson background \cite{Sfetsos:2010uq}.
This exceptional case was expected to be singled-out by our approach in light of the results of \cite{Borsato:2017qsx}. Indeed, our string solitons shows no sign of non-integrability. For any deformation away from that background or any other background dual to a generic quiver CFT, we have proven that they will show analytic signs of non-integrability in the Liouville sense.

We have complemented our study with a numerical analysis where the characteristic chaotic indicators were calculated. In fact, we found that for deformations away from the Sfetsos-Thompson solution, the Lyapunov exponent is nonzero. For more generic quiver CFTs we computed the Poincare sections and power spectra, that also display signs of chaotic (hence non-integrable) dynamics.

In the paper  \cite{Gadde:2012rv} the authors showed that
${\cal N}=2$ SQCD, that is the theory with one gauge group $SU(N_c)$
and $N_f=2N_c$, is {\it not} integrable. They considered
operators of the form
\begin{equation}\label{eq:op}
{\cal O}=Tr(\phi^{k_1} M^{k_2} \phi^{k_3} M^{k_4}....), 
\end{equation}
where $\phi$ is some adjoint operator made out of fields
in the vector multiplet and $M$ is a dimer operator in the adjoint,
constructed out
of fundamental field with the flavour index contracted. Calculating two body S-matrix scattering, they showed that
these dimer fields play an important role in the absence of integrability.

The conjecture that there may be a subsector of the theory that preserves
integrability was proposed in \cite{Gadde:2012rv} and made more concrete
in \cite{Pomoni:2013poa}. In fact, when only fields belonging  to the
vector multiplet are considered, the integrable structure is inherited from that of ${\cal N}=4$ SYM, up to a rescaling of the gauge coupling.

These perturbative results complement the findings of our study, that is
non-perturbative in nature. We suggest that for quiver theories with
fundamental fields, the presence of the flavour group is at the root
of the non-integrability we found. Indeed, when our strings explore
the regions close to the end of the quiver our numerical study
shows the KAM tori becoming diffuse. More analytically, a deviation
from the Sfetsos-Thompson solution (by the addition of a flavour group),
also displayed non-integrability. Also, the conjecture that the sub-class 
of operators not containing dimer fields
inherit their integrable properties from ${\cal N}=4$ SYM is nicely mirrored in our approach. In fact,  strings uncharged under $SU(2)_R$---with $k=0$ in eq.(\ref{confix}), but charged under $U(1)_r$ (with nonzero $\lambda$), are dual to an long operator made out of vector multiplet fields only. The analysis shows integrability of those solitons.

In summary, this work belongs to a line of studies of  ${\cal N}=2$
SCFTs and holographically shows that generically they are not-integrable.   The paper opens some interesting topics for exploration. We list a few of these below and hope to return to them in future work.
\begin{itemize}
\item{It would be interesting to understand in more detail the holographic background we associated with the deformation away from the Sfetsos-Thompson solution---see around eq.(\ref{vvvccc}). The calculation of some observables, charges and structure of singularities in terms of the deformation parameter should give clues about the dual CFT.}


\item{It seems natural to extend our study to CFTs with similar characteristics. Namely to holographic backgrounds with AdS-factors and some $SU(2)$ isometry, preserving eight Poincare supercharges. Some examples come to mind. The Gaiotto-Witten CFTs with holographic dual summarised  in \cite{Assel:2011xz} and the recently discussed five dimensional CFTs based on intersections of D5-NS5-D7 branes summarised in \cite{DHoker:2017zwj}. Notice that both systems admit a background obtained via non-Abelian T-duality (hence analog to Sfetsos-Thompson), given in the works \cite{Lozano:2016wrs}  and  \cite{Lozano:2012au} respectively. }
\item{The zero-winding case---see eq.(\ref{eqdiffxx}) for the case $k=0$--- becomes integrable in the Liouville sense. This translates to the integrability of this particular geodesic motion. It seems plausible, following the ideas of \cite{Chervonyi:2013eja}, to study generically the integrability of  geodesics in Gaiotto-Maldacena backgrounds.  }

\item{Finally and more interestingly, it would be nice to clearly establish what characteristic of the Gaiotto CFT is introducing the non-integrability and/or the chaotic behaviour. One intuitive answer indicates that is the presence of fundamental fields (flavours) that is  producing Poincare sections with signs of chaos.  One should like a more precise identification of the corresponding operators, like that in eq.~\eqref{eq:op} that drive this feature.}

\end{itemize}
We hope to report on these issues in the future.

\section{Acknowledgements}
C.N. is Wolfson Fellow of the Royal Society. DCT is supported by a Royal Society University Research Fellowship {\em Generalised Dualities in String Theory and Holography} URF 150185.  DR is supported through the Newton-Bhahba Fund and he would like to acknowledge the Royal Society UK and the Science and Engineering Research Board India (SERB) for financial assistance. We have been partly supported by 
by STFC grant ST/P00055X/1.
We would like to thank  Prem Kumar,  Leo Pando Zayas,  Jeroen Van Gorsel and  Konstantin Zarembo for enlightening discussions on the topic of this paper.

\newpage
\appendix

\section{Kovacic's algorithm and Rudiments of Differential Galois Theory }
\label{sec:Tech}

In this article we need to establish if certain linear ordinary differential equations are integrable in the Liouvillian sense. That is to say we wish to know if a differential equation is ``symmetric'' enough to admit   solutions given in closed form in terms of a finite composition of elementary functions.   The symmetries here are linear transformations of the space of solutions that respect both algebraic and differential relations between solutions.   These symmetries are described by Picard-Vessiot or differential-Galois theory.  Since this is a concept less familiar amongst physicists we give a brief orientation below.\footnote{Our presentation is at a rather telegraphic and superficial level and we recommend the reader to the books   \cite{Ruiz:1999,Singer1,Singer2}.  The articles \cite{ziglin1982,Morales1994140,Ruiz:1999,MoralesRuizRamis2001,MoralesRuiz2007845}   explain the application of these techniques to Hamiltonian systems.} 

We will describe two different types of arguments. One is based on usual lore and manipulations with  differential equations,
the other on Lie-group theoretical arguments (Picard-Vessiot theory). Both developments are familiar to mathematicians and were used by Kovacic in his work 
\cite{Kovacic85}

\subsection{The differential equation approach}
Here, we briefly describe Kovacic's algorithm \cite{Kovacic85}.
Let us start considering a linear differential equation,
\begin{equation}
y''(x)+ {\cal B}(x) y'(x) +{\cal A}(x) y(x)=0,
\label{kov1}
\end{equation}
where ${\cal A}(x),{\cal B}(x)$ are complex rational functions. We are concerned with the existence of  solutions that can be expressed
in terms of algebraic functions, exponentials, trigonometric and integrals of the above. These are  called  `Liouvillian' solutions.

 The algorithm
of \cite{Kovacic85} provides such solutions or shows that there are none (in such case we refer to the equation (\ref{kov1}) as non-integrable).
We will not describe the algorithm in itself, as it is already implemented in different softwares. 
We limit ourselves to explain the `logic' behind the derivation and some {\it necessary but not-sufficient} conditions 
that a combination the functions ${\cal A},{\cal B}, {\cal B}'$ must satisfy, for the eq.(\ref{kov1}) to be Liouville integrable.

We start by rewriting  the differential equation as,
\begin{eqnarray}
& & y(x)= e^{\int w(x)-\frac{{\cal B}(x)}{2} dx},\nonumber\\
& & w'(x) +w(x)^2=V(x)=\frac{2{\cal B}' + {\cal B}^2-4 {\cal A}}{4}.\label{kov2}
\end{eqnarray}
It was shown that if the function $w(x)$ is algebraic of degrees 1,2,4,6, or 12, then the eq.(\ref{kov1}) is Liouville integrable \cite{Kovacic85}.
This results comes from the application of Galois theory to differential equations (which is Piccard-Vessiot theory). As we shall discuss below,
this formalism studies the most general group of invariances of the differential equation (\ref{kov1}), the transformations that act on the solutions of the equation,  that is a subgroup of $SL(2,C)$. Kovacic \cite{Kovacic85} showed that there are four possible cases of subgroups of $SL(2,C)$ that can arise
\begin{itemize}
\item{{\bf Case 1:} the subgroup is generated by the matrix of the form   
\[
   G=
  \left[ {\begin{array}{cc}
   a & 0 \\
   b & \frac{1}{a} \\
  \end{array} } \right],
\]
with $a,b$ complex numbers. In this case $w(x)$ is a rational function of degree 1.}

\item{{\bf Case 2:} the subgroup of $SL(2,C)$ is generated by matrices of the form
\[
   G=
  \left[ {\begin{array}{cc}
   c & 0 \\
   0 & \frac{1}{c} \\
  \end{array} } \right],\;\;\;
   G=
  \left[ {\begin{array}{cc}
   0 & c \\
   -\frac{1}{c} & 0 \\
  \end{array} } \right],
\]
in this case the functions $w(x)$ is rational of degree 2
}
\item{{\bf Case 3:} the situation in which $G$ is a finite group, not included in the two above cases. In such case, the degree of $w(x)$ is either 4,6 or 12.}
\item{{\bf Case 4:} the group is $SL(2,C)$ and the solution for $w(x)$, if they exist are non-Liouvillian.}
\end{itemize}
Kovacic provided not only an algorithm to find the solutions in the first three cases above, but also
 a set of {\it necessary but not sufficient} conditions that the function $V(x)$ in eq.(\ref{kov2}) must satisfy to be in any of the first three cases detailed above
 \cite{Kovacic85}. For each of the cases as ordered above, the conditions are :
\begin{itemize}
\item{{\bf Case 1:} every pole of $V(x)$ has order 1 or  has even order. The order of the function $V(x)$ at infinity\footnote{The order of a rational function at infinity is the highest power of the denominator minus the highest power of the numerator.}  is either even or greater than 2.}
\item{{\bf Case 2:} $V(x)$ has either one pole of order 2, or  poles of odd-order greater than 2 . }
\item{{\bf Case 3:} the order of the poles of  $V$ does not exceed 2, and the order of $V$ at infinity is at least 2.}
\end{itemize}
If none of the above is satisfied, the analytic  solution (if it exists), is non-Liouvillian.
In Section \ref{manaza}, we shall explore some examples in the main part of this paper in light of these statements. Before 
that, let us discuss the group theoretical Picard-Vessiot viewpoint of the ideas in this section.

\subsection{The group theoretical approach}
Consider a homogenous $n^{th}$ order linear differential equation  $L(y) = y^{(n) } + a_{n-1} y^{(n-1)} + \dots + a_0 y = 0$ with coefficients in some differential field $k$ with constants $\mathbb{C}$.  The $n$-independent solutions of $L(y)$ form a vector space $V$ over  $\mathbb{C}$.  A differential field $K$ is said to be a Picard-Vessiot (PV) extension  of $k$ for  $L$ if $K$ is generated over $k$ by the solutions of $L$.  A PV extension--whose existence and uniqueness was shown by Kolchin-- is thus the smallest such extension of $k$ that contains the  $n$-independent solutions of $L(y)$.  The differential Galois group $G= \partial\textrm{Gal}(K/k)$ is the group of automorphisms of the PV extension  $K$  that commute with the derivative and which leave elements of $k$ fixed.   The condition that $L(y)$ be Liouvillian integrable is now more formally stated as demanding that $G^0$, the identity connected component of the  differential Galois group is solvable.  If $G^0$ is Abelian then  $L(y)=0$ is integrable.

To make this rather more concrete let us do a trivial example.  $y''(x) + \frac{1}{x}y'(x)=0$   is an integrable equation with a solution space 
   $V =  1 \oplus \log(x)$.    Since $1$ is just a constant of the base field any $\sigma \in   G$  should obey $\sigma(1) = 1$.  For the action of $\sigma$ on $\log(x)$  we have $\partial \sigma(\log(x)) =  \sigma(\partial  \log(x)) = \sigma(\frac{1}{x} ) = \frac{1}{x}$ in which we used that $\sigma$ commutes with derivation and that $\frac{1}{x}$ is by definition left invariant by $\sigma$.   We can then extract $\sigma(\log(x))$ by integration to find  
   \be
   \sigma : (1 , \log(x) )  \mapsto  (1 , \log(x) )\left( \begin{array}{cc}1 & c \\ 0 & 1 \end{array} \right) \ , \quad  c \in \mathbb{C}  \ , 
   \ee
   and so that the Galois group is simply the Abelian additive group on $\mathbb{C}$ formed by the composition of   such $\sigma$. This is an example of Case 1 in the preceding discussion.  In general, unlike in this example, the construction of differential Galois groups is not facile, however can be achieved algorithmically (for second order equations one has an algorithm due to Kovacic and in generality from Hrushovski).  
   
 The concept of differential Galois group is at first sight similar to that of monodromy.    Transporting the Wronskian matrix of fundamental solutions  ${\bf Z}$ around some closed path $\gamma$ avoiding singularities of $L(y)$ generates some new fundamental solution matrix  ${\bf Z}_\gamma = {\bf Z} M_\gamma$.   Providing that the singularities  of $L$ are regular (i.e. $a_i$ has a pole of at most order $n-i$ at singular points ) then monodromy is dense in the Galois group.   However in general singularities need not be regular.   In this case a theorem by Ramis establishes the Galois group as being generated by formal monodromy that comes from $x \to x e^{2\pi i}$,  the so-called Ramis/exponential torus (a subgroup of $(\mathbb{C}^\ast)^n$ whose precise definition we shall not need) and the Stokes multipliers.    
 
 This is well exemplified by the Bessel equation,
 \be \label{eq:Bessel}
 x^2 y''(x) + x y'(x)  +(x^2 - n^2 ) y = 0  \ , 
 \ee
 in which $x=0$ is a regular point and $x=\infty$ is irregular.  To simplify matters let us just consider $\nu$ not an integer for which $J_{+\nu}(x)$ and $J_{-\nu} (x)$ are independent solutions.  Around $x=0$ these are described by convergent series  $J_\nu(x) =  \frac{1}{\Gamma(1+\nu) }\left( \frac{x}{2}\right)^\nu \left(1 - \frac{1}{1+\nu} \left( \frac{x}{2}\right)^2 + O(x^4)  \right) $ with monodromy $J_\nu ( e^{2\pi i }x )= e^{2 \pi i \nu}  J_\nu (  x)$.   Hence the monodromy around the origin is simply $M_0 = \diag( e^{2\pi i \nu} ,e^{-2\pi i \nu} )$.   On the other hand expansions around  $x=\infty$ are asymptotic.  Translating into first order equations by introducing the vector ${\bf Y}(x)=(y(x) , y'(x) )$  and letting $x= z^{-1}$ we have 
 \be
 d_z {\bf Y} =  {\bf A}    \cdot {\bf Y}  \ , \quad  {\bf A}= \left( \begin{array}{cc} 0 & 1 \\ - \frac{1}{z} & -\frac{1}{z^4} + \frac{\nu^2}{z^2}  \end{array} \right) \ . 
 \ee 
 The formal fundamental solution matrix ${\bf Z}$ which obeys the same equation can be factored as 
 \be
 {\bf Z} =\hat{\phi}(z) z^{{\bf L}} e^{Q(z)}  \ , 
 \ee 
 in which $\hat{\phi}(z)$ is a matrix consisting of formal (i.e. asymptotic) power-series $\sum_{0}^\infty a_n z^n$ and in the case at hand ${\bf L}=  \frac{1}{2} \mathbf{1}$ and $Q= \diag(q_1,q_2) = \diag(\frac{i}{z},  \frac{-i}{z})$.      The {\em formal} monodromy at infinity is sensitive only to the square root in $z^L$ and so $\hat{M}_\infty = - {\bf 1}$.   Within a given sector the formal $\hat\phi(z)$ can be Borel resumed however as the phase of $z$ is varied the result of this procedure changes exactly as one crosses singular directions (anti-stokes rays).  These occur at in a direction  $arg(z)= \theta$ for which $Re( q_1 - q_2 ) = 0 $. It is in these directions  that dominant and sub-dominant asymptotic behaviours switch roles.  The summation before and after crossing the singular direction specified by $arg(z)= \theta$  must then be related via $\phi_+(z) = \phi_-(z) {\cal S}_\theta$ where  ${\cal S}_\theta$    called a Stokes multiplier.   For the Bessel function the singular directions are $\theta = \frac{\pi}{2} , \frac{3 \pi}{2}$ and corresponding multipliers are 
 \be
 {\cal S}_{\frac{\pi}{2} } =   \left( \begin{array}{cc} 1 & \lambda  \\  0 &  1  \end{array} \right) \ , \quad  {\cal S}_{\frac{3\pi}{2} } =   \left( \begin{array}{cc} 1 & 0 \\  -\mu   &  1  \end{array} \right) \ .
 \ee
We don't need to evaluate the mulitpliers directly, instead we follow the nice argument in  \cite{Ruiz:1999} to relate the actual monodromy at the origin to the formal monodromy at $\infty$ via  $M_0 =  {\cal S}_{\frac{\pi}{2} } \cdot  \hat{M}_\infty   \cdot  {\cal S}_{\frac{3\pi}{2} } $ and by taking a trace of this relation we constrain the Stokes multipliers to obey
\be
\lambda \mu  = 4 \cos^2 (\pi n ) \ . 
\ee
If $n \notin \mathbb{Z}+ \frac{1}{2} $ the Stokes multipliers are not-vanishing and  $ {\cal S}_{\frac{\pi}{2} },  {\cal S}_{\frac{3\pi}{2} } $ do not commute; hence   $G^0$ is non-Abelian and the Bessel equation is not Liouvillian integrable.  On the other-hand when  $n  \in \mathbb{Z}+ \frac{1}{2} $ the  formal series in  $\hat{\phi}(z)$ actually terminate and the Stokes multipliers vanish and the Bessel equation becomes integrable.  Indeed we have that e.g.  
\begin{equation}
J_{\frac{1}{2} }(x)  = \left(  \frac{2 }{\pi x} \right)^\frac{1}{2} \sin x  \ . 
\end{equation}

\subsection{Some  examples}\label{manaza}
Along the lines of the discussion above, let us study the   function $V(x)$  in eq.(\ref{kov2}) for some of the examples in this paper. We start with the Sfetsos-Thompson solution. The associated NVE differential equation was discussed in eq.(\ref{natdliouville}). The coefficientes ${\cal A},{\cal B}$ read in this case (we take $k=E=1$ to avoid a cluttered notation),
\begin{equation}
{\cal A}= 1+ 2 \frac{(4 \tau^2 +3)}{4\tau^2+1},\;\;\; {\cal B}=\frac{2}{4\tau^3+\tau}.
\end{equation}
The coefficients are rational, so we construct the potential function
\begin{equation}
4 V(\tau)= 4{\cal A}^2  -{\cal B}^2-2 {\cal B}'=3+ \frac{12}{(4 \tau^2+1)^2} +\frac{4}{4\tau^2+1} +\frac{\gamma}{\tau^2}.
\end{equation}
The last term  has a coefficient $\gamma= E(E-1)$, that vanish for our particular choice of constants. 
Note that his function is the one appearing in eq.(\ref{dadaya}) and after a rescaling and choice of constants, in eq.(\ref{estaxxx}). Analysing the $V$-function, we see that it satisfies all the three possible necessary conditions listed above. Hence, if there is a Liouvillian solution, Kovacic's algorithm should find it. Indeed, this happens when feeding the differential equation to any software, the (Liouvillian) solution in 
eq.(\ref{liouvillianNATD}) is found.

We now move to study the case of the deformed Sfetsos-Thompson solution in eqs(\ref{vdef})-(\ref{coeffdef}). For this analysis to be  simpler, we shall consider 
the deformation parameter $\epsilon\sim 0$ and keep only up to linear order in a series expansion in  $\epsilon$. In this case the coefficients  in eq.(\ref{coeffdef}) are
\begin{equation}
{\cal A}= \frac{(12 \tau^2 +7)}{4\tau^2+1} +\epsilon \left(\frac{5\tau+8\tau^3 +16\tau^5}{2(4\tau^2+1)^2}\right),\;\;\; {\cal B}=\frac{2}{4\tau^3+\tau} +\epsilon\left( \frac{1-16\tau^2 -16\tau^4}{4(4\tau^2+1)^2}   \right).
\end{equation}
The associated potential function is,
\begin{eqnarray}
 & &V=\frac{19 + 40 \tau^2 +48\tau^4}{(4\tau^2+1)^2}+\frac{\epsilon}{4\tau (4\tau^2+1)^3}\left(128\tau^8 +96\tau^6 +40\tau^4 +50\tau^2 -1  \right).
\end{eqnarray}
We can check that the three criteria are failed. So the solution to the NVE equation is non-Liouvillian.

%

Finally, let us analyse the case of the Maldacena-N\'u\~nez solution in Section \ref{sectionmnxx}.   In this case  $V_{MN}$ solution of the Laplace equation dictates the functions   ${\cal A},{\cal B}$  for $0\leq \eta\leq N$ to be
\begin{eqnarray}
& & {\cal A}=k^2+ \sqrt{2} E k \frac{(3 N^2 +\eta^2)}{(N^2+\eta^2)\sqrt{|\eta^2-N^2|}}\nonumber\\
& & {\cal B}=E\frac{(\eta^4+ 3 N^2 \eta^2 -2 N^4)}{\eta (\eta^4-N^4)}. \end{eqnarray}
For $N\leq \eta$ we find,
\begin{eqnarray}
& & {\cal A}=k^2+ \sqrt{2} E k \frac{1}{\sqrt{|\eta^2-N^2|}}\nonumber\\
& & {\cal B}=E\frac{\eta}{ (\eta^2-N^2)}.\nonumber
\end{eqnarray}
Here we understand that  $\eta= E\tau$.   We simplify the analysis by  choosing again all arbitrary constants $E=k=N=1$ and considering the system in the interval $[0,1]$ by choosing appropriate initial conditions.   Thus
\begin{equation}
{\cal A}= 1+ \sqrt{2} \frac{( \tau^2 +3)}{(\tau^2+1)\sqrt{1-\tau^2}},\;\;\; {\cal B}=-\frac{\tau^4+3\tau^2-2}{-\tau^5+\tau}.
\end{equation}
In this case, the coefficients are not rational. We should then change variables as,
\begin{equation}
\tau=\sqrt{1-z^2},\;\; \frac{dz}{d\tau}=-\frac{\sqrt{1-z^2}}{z},\;\; \frac{d}{dz}(\frac{dz}{d\tau})= \frac{1}{z^2\sqrt{1-z^2}},\;\;\;
\frac{d}{dz}\frac{dz}{d\tau}=\frac{1}{z^2\sqrt{1-z^2}}
\label{changetauz}
\end{equation}
The derivatives change according to,
\begin{eqnarray}
& &\dot{x}=\frac{dx}{d\tau}= \frac{dx}{dz} \frac{dz}{d\tau}= -x'(z)\frac{\sqrt{1-z^2}}{z},\nonumber\\
& & \ddot{x}=x'' (\frac{dz}{d\tau})^2 + x' \frac{d}{dz}(\frac{dz}{d\tau})\times \frac{dz}{d\tau}.\nonumber
\end{eqnarray}
The NVE equation changes according to,
\begin{eqnarray}
& & \ddot{x}(\tau)+ {\cal B}\dot{x}(\tau)+{\cal A} x(\tau)=0\to x''(z) +{\cal C} x'(z) + {\cal D} x(z)=0,\nonumber\\
& & {\cal C}= \frac{{\cal B} + \frac{d}{dz}(\frac{dz}{d\tau}) }{\frac{dz}{d\tau}}=-z \frac{{\cal B}(z) +\frac{1}{z^2\sqrt{1-z^2}}}{\sqrt{1-z^2}}=\frac{z^3-4 z}{z^4-3z^2+2},\nonumber\\
& & {\cal D}=\frac{{\cal A}}{(\frac{dz}{d\tau})^2}= \frac{z^2}{1-z^2}{\cal A}(z)=-\frac{\sqrt{2}z (z^2-4) + z^4 -2 z^2}{z^4-3 z^2+2}.
\end{eqnarray}
Plugging this in the  function $-4V(z)=2{\cal C}' +{\cal C}^2 -4{\cal D} $, we find
\begin{eqnarray}
& & V(z)=4+\frac{10-z^2}{(z^4 -3 z^2+2)^2} +\frac{4\sqrt{2} z^3 +5z^2 -16\sqrt{2} z-13}{z^4-3z^2 +2}
\end{eqnarray}
In this case we see that the three necessary criteria are failed. The solution to the NVE is non-Liouvillian.

\section{Non-integrability in generic Gaiotto-Maldacena Backgrounds}\label{appendixgm}
We will consider a generic Gaiotto-Maldacena background. All 
the elements of the metric and other fields can be written in 
terms of the function $\dot{V}(\sigma, \eta)$ and its derivatives.

Generically, the potential function and its derivatives are \cite{ReidEdwards:2010qs},
 \cite{Aharony:2012tz},
\begin{eqnarray}
& & \dot{V}(\sigma,\eta)=\sum_{n=1}^{\infty}A_n (\omega_n\sigma) 
K_1(\omega_n \sigma)\sin(\omega_n \eta),\;\;\; 
\dot{V}'(\sigma,\eta)=\sum_{n=1}^{\infty}A_n \omega_n (\omega_n\sigma) 
K_1(\omega_n \sigma)\cos(\omega_n \eta),\nonumber\\
& & \ddot{V}(\sigma,\eta)=\sum_{n=1}^{\infty}A_n (\omega_n\sigma)^2 
K_0(\omega_n \sigma)\sin(\omega_n \eta),\;\; 
{V''}(\sigma,\eta)=-\sum_{n=1}^{\infty}A_n \omega_n^2 
K_0(\omega_n \sigma)\sin(\omega_n \eta).\nonumber
\end{eqnarray}
Here, we used that $\omega_n= \frac{n\pi }{N_5}$ and that for 
$\sigma=0$ we have,
\be
\tilde{\lambda}(\eta)=\dot{V}(0,\eta)=\sum_{n=1}^{\infty}A_n \sin(\omega_n \eta).\nonumber
\ee
In other words, the coefficients $A_n$ are the Fourier 
decomposition of $\tilde{\lambda}(\eta)$.

The authors of the paper
\cite{Aharony:2012tz} proposed an expansion 
close to $\sigma=0$ for a value of $\eta=\eta_i$ 
where a change in slopes is found. They set 
\begin{eqnarray}
\sigma= r \cos\theta,\;\;\;\; \eta=\eta_i+ r \sin\theta,\nonumber
\end{eqnarray}
and found the expression for the metric and background fields 
close to that point---see Section 4.2.2 in 
\cite{Aharony:2012tz}. 
The fields  relevant to our purposes are,
\begin{eqnarray}
& & ds^2= \sqrt{g(r)}\left[4 AdS_5 +d\chi^2+\sin^2\chi d\xi^2  \right] 
+\frac{1}{\sqrt{g(r)}}\left[dr^2+ r^2d\theta^2+ r^2 \sin^2\theta d\beta^2  
\right],\nonumber\\
& & B_2= -2\eta_i \sin\chi d\chi\wedge d\xi.
\end{eqnarray}
The function $g(r)= 4\frac{\lambda(\eta_i)}{(a_{i-1}-a_i)} r$.

As in previous sections, we propose a configuration of the form,
\begin{eqnarray}
 t=t(\tau),\;\;\; r=r(\tau),\;\;\; 
\theta=\theta(\tau),\;\;\; \chi=\chi(\tau),\;\;\xi= k\sigma,\;\; 
\beta=\nu\sigma.\nonumber
\end{eqnarray}
The effective Lagrangian is,
\begin{eqnarray}
L_{eff}= \sqrt{g}\left[ 4\dot{t}^2 +k^2\sin^2\chi- \dot{\chi}^2 \right]+
\frac{1}{\sqrt{g}}\left[-\dot{r}^2 -r^2\dot{\theta}^2 +\nu^2r^2\sin^2\theta  
\right] +4 k\eta_i\sin\chi \dot{\chi}.
\end{eqnarray}
Notice that the last term, the one induced by the B-field 
is a total derivative, as expected, since the B-field turns out 
to be pure gauge in the expansion. 

We calculate the equations of motion. We find the usual conservation equation,
$\dot{t}=\frac{E}{ \sqrt{g(r)} }$. 
Replacing this in the other equations, we have,
\begin{eqnarray}
& & 4 g(r)  \ddot{r}= g'(r)(\dot{r}^2 +4 E^2)+\left(4 r g(r)- r^2 g'(r)\right)
(\dot{\theta}^2-\nu^2\sin^2\theta)+g(r) g'(r) (\dot{\chi}^2 - k^2\sin^2\chi).\nonumber\\
& & 2 g(r) r \ddot{\theta}= -\nu^2 g(r) r \sin(2\theta) -
(r g'(r)-4g(r))\dot{r}\dot{\theta},\nonumber\\
& & 2 g(r)\ddot{\chi}= - k^2 g(r)\sin(2\chi)-g'(r)\dot{r} \dot{\chi}.
\label{eqslag}
\end{eqnarray}
We observe that for $\theta=\dot{\theta}=\ddot{\theta}=0$ 
and $\chi=\dot{\chi}=\ddot{\chi}=0$,
we solve automatically the $\theta$ and $\chi$-equations in (\ref{eqslag}).
If we replace this in the $r$-equation we find,
\be
4 g(r)\ddot{r}- g'(r)(4E^2+\dot{r}^2)=0.
\label{xxcc}
\ee
This equation has a complicated solution in terms of 
exponential, cubic roots, etc. We will not write it here.

A small fluctuation around the $\chi=0+\epsilon x(\tau)$
and $\theta=0+\epsilon y(\tau)$, gives at leading order in 
$\epsilon$ the equations,
\begin{eqnarray}
& &  \ddot{x} +\frac{g'(r)\dot{r}}{2g(r)} \dot{x} +k^2 x=0,\nonumber\\
& &  \ddot{y} +\frac{(2 g(r)\dot{r}-g'(r) r \dot{r})}{ g(r) r} 
\dot{y} +{\nu^2} x=0.
\label{xxccbb}
\end{eqnarray}
One can calculate the time-dependent coefficients of the 'friction terms'
and attempt to solve the complicated equations.
This leads to solutions involving  
a complicated combination of special functions. 

In the very simple case in which the integration constant $E=0$, the solution of eq.(\ref{xxcc}) 
is $r=t^{4/3}$. Then eq.(\ref{xxccbb}) leads to a simple 
solution in terms of Bessel functions. These are not Liouvillian solutions, hence
the string soliton is non-integrable. 

In summary, we find non-integrability in 
Gaiotto Maldacena backgrounds, with
the string soliton at a generic kink-point.

   \section{Integrability in the Non-Abelian T-dual of $AdS_5\times S^5$}
   \label{sec:S5int}
   
  The preservation of integrability under non-Abelian T-duality of various (super)-coset string sigma-models was shown in   \cite{Borsato:2016pas,Borsato:2017qsx}. In this appendix we reinforce these results by giving a very explicit treatment of the integrability of the bosonic theory with an $S^5$ target space dualised with respect to an $SU(2)_L$ isometry acting inside the sphere.  Our considerations apply equally well to the bosonic sector of the $AdS_5 \times S^5$  string since the $AdS_5$ space is   decoupled from the five sphere  (except via Virasoro constraints).   
   
 We view the $S^5$ as a coset $SO(6) / SO(5)$ and introduce an explicit representation for the algebra  $\frak{g} = \frak{su}(4) \cong \frak{so}(6)$.    We follow the conventions of \cite{Arutyunov:2009ga}   first define  4d Dirac matrices\footnote{We use
 \be
\begin{aligned} \gamma_1 = \left(
\begin{array}{cccc}
 0 & 0 & 0 & -1 \\
 0 & 0 & 1 & 0 \\
 0 & 1 & 0 & 0 \\
 -1 & 0 & 0 & 0 \\
\end{array}
\right) \ , \quad 
\gamma_2 = \left(
\begin{array}{cccc}
 0 & 0 & 0 & i \\
 0 & 0 & i & 0 \\
 0 & -i & 0 & 0 \\
 -i & 0 & 0 & 0 \\
\end{array}
\right)\ ,   
\gamma_3= \left(
\begin{array}{cccc}
 0 & 0 & 1 & 0 \\
 0 & 0 & 0 & 1 \\
 1 & 0 & 0 & 0 \\
 0 & 1 & 0 & 0 \\
\end{array}
\right)\ , \quad 
\gamma_4 = \left(
\begin{array}{cccc}
 0 & 0 & -i & 0 \\
 0 & 0 & 0 & i \\
 i & 0 & 0 & 0 \\
 0 & -i & 0 & 0 \\
\end{array}
\right) \ , 
\end{aligned}  \nonumber 
 \ee}
 which obey the Clifford algebra $\{ \gamma_i,  \gamma_j \} = 2\delta_{ij} $ and  supplement them with $\gamma_5 = - \gamma_1 \gamma_2 \gamma_3 \gamma_4$.    Generators of $\frak{so}(5) \in \frak{so}(6)$  are given by $n_{ij} = - n_{ji} = \frac{1}{4} [\gamma_i , \gamma_j]$ and the remaining five coset generators are given by $n_{i 6} = \frac{i}{2} \gamma_i$.   The  $\frak{su}(4)$ algebra is $\mathbb{Z}_2$ graded by 
\be
\Omega(\frak{g} ) = K \cdot \frak{g} \cdot K^{-1} \ , \quad K = - \gamma_2 \gamma_4 \ . 
\ee
Denoting  $\frak{g}^{(k)} = \{ X \in \frak{g} | \Omega(X) = i^k X \} $ we have $\frak{g}=\frak{g}^{ (0)} + \frak{g}^{ (2)}$ with $\frak{g}^{ (0)}$ being the $\frak{so}(5)$ subgroup and $\frak{g}^{(2)}$ the coset generators.    Notice that this is a symmetric coset since $[\frak{g}^{ (0)},\frak{g}^{ (0)}] \in \frak{g}^{ (0)}$, $[\frak{g}^{ (2)},\frak{g}^{ (2)}] \in \frak{g}^{ (0)}$ and $[\frak{g}^{ (0)},\frak{g}^{ (2)}] \in \frak{g}^{ (2)}$.  We let $P^{(k)}$ be the projector onto $\frak{g}^{(k)}$ .

To define the sigma model on the $S^5$  we introduce a coset representative $G$ from which we construct an algebra valued one-form 
\be
\frak{a} = - G^{-1} d G = \frak{a}^{(0)} +  \frak{a}^{(2 )}\  .  
\ee
The Lagrangian is then given by the pull back    
\be\label{eq:Lagcoset} 
{\cal L } = Tr (    \frak{a}^{(2 )}_+ ~  \frak{a}^{(2 )}_-   )    \ , 
\ee
where $\sigma^\pm = \tau \pm \tilde\sigma$ are world sheet light cone coordinates.  The equations of motion and Bianchi identities of this theory are encapsulated by a Lax connection with light cone components  
\be\label{eq:LaxCoset}
\frak{L}_\pm = \frak{a}^{(0)}_\pm +  (u_1 \mp u_2 )  \frak{a}^{(2)}_\pm  \ , \quad u_1^2 - u_2^2 = 1 \ , 
\ee
and which obeys the flatness condition 
\be
[\partial_+  +\frak{L}_+ , \partial_- +  \frak{L}_-]=0 \ . 
\ee
 
 Various options are available for the parametrisation of the coset and to proceed explicitly we must pick one.  We will make a choice that will most easily allow us to perform the T-dualisation of the sigma model with respect to an $SU(2)$  generated by 
 \be
 T_1 = -\frac{1}{2}( n_{12}  + n_{34} ) \ , \quad T_2 = -\frac{1}{2}( n_{13}  - n_{24} ) \ , \quad T_3 = -\frac{1}{2}( n_{14}  + n_{23}  ) \ ,  \quad [T_a, T_b] = \epsilon_{abc} T_c  \ .   
 \ee
 We will then mimic the standard Euler angles and define
 \be
 G = H  \hat{G}   \ , \quad H = \exp[ \phi T_3] \cdot \exp[ \theta T_2] \cdot \exp[ \psi T_3]  \  , \quad \hat{G} = \exp[ \frac{i}{2} \tilde{\phi}  \gamma_5 ]\cdot  \frac{1}{\sqrt{1+r^2}}  \left( 1 + i r \gamma_3 \right)  \ . 
 \ee
The target space metric of the Lagrangian eq.~\eqref{eq:Lagcoset} reads 
\be
ds^2 = d\alpha^2 + \sin^2\alpha  d\tilde{\phi}^2  + \frac{1}{4} \cos^2 \alpha \left( l_1^2 + l_2^2 + l_3^2 \right) 
\ee
in which we use the $SU(2)$ left invariant one-forms
 \be
 l_1=\sin  \theta   \cos  \psi   d\phi -d\theta  \sin  \psi  \ , \quad l_2 = \cos \psi  (\sin \theta   \tan  \psi  d\phi +d\theta ) \ , \quad l_3 = \cos  \theta  d\phi +d\psi \ , 
 \ee
 and the angle $\sin \alpha = \frac{1-r^2}{1+r^2} $.   The virtue of this parametrisation is that we have chosen a gauge for the Lax connection that makes the $SU(2)$ isometry manifest.  Explicitly we find that 
 \be\label{eq:a0}
  \frak{a}^{(0)} = \cos \alpha\, d\tilde{\phi}\, n_{35} + \frac{1}{2} l_1 ( n_{12} + \sin \alpha\, n_{34} ) + \frac{1}{2} l_2 (- n_{24} + \sin \alpha\, n_{13} )  +  \frac{1}{2} l_3 ( n_{14} + \sin \alpha\, n_{23} ) 
 \ee
 and 
 
  \be\label{eq:a2}
  \frak{a}^{(2)} =  - d\alpha \, n_{36} + \sin \alpha  d\tilde{\phi}\, n_{56} - \frac{1}{2} \cos \alpha  \left[  l_1 \, n_{46} +l_2 \, n_{16} + l_3 \, n_{26}\right] \ . 
 \ee
Notice our choice is such that the Euler angles only appear in their left invariant one-form combinations, this is vital because the T-duality rules are local when acting on these one-forms but non-local when applied to the individual coordinates $(\theta, \phi, \psi)$.  

We now proceed to the T-dualisation of the Lagrangian eq.~\eqref{eq:Lagcoset} with respect to the $SU(2)_L$ symmetry for which the $l_i$ are invariant one-forms.   Some useful quantities are 
\be
\hat{J}=  -\hat{G}^{-1} d\hat{G}    \ ,  \quad T_a^{\hat{G}}  = \hat{G}^{-1} T_a \hat{G}     \ ,  \quad 
G_{ab} = - Tr (T_a^{\hat{G}}P^{(2)}  T_b^{\hat{G}} ) \ , \quad Q_a =  Tr( \hat{J} P^{(2)}  T_a^{\hat{G}}) \ .
\ee 
In the case at hand $G_{ab} = \frac{\cos^2 \alpha }{4} \delta_{ab}$ and $Q_a = 0$.  
Then the T-dual theory is given by the Lagrangian 
\be\label{eq:Lagcosetdual} 
{\cal L }_{dual}  = -Tr (  \hat{J}_+P^{(2)}   \hat{J}_-   )    + \partial_+ v_a (M^{-1})^{ab} \partial_- v_b  \ ,  
\ee
 with 
 \be
 M_{ab} = G_{ab} + \epsilon_{abc} v_c  \ . 
 \ee
 With the transformation 
  \be
v_1 =   \frac{1}{4} r \cos \theta  \ , \quad v_2 =   \frac{1}{4}  r  \sin \theta \cos \phi \ , \quad v_3 =  \frac{1}{4} r \sin \theta \sin \phi  \ , 
\ee
the T-dual metric is given by 
\be
ds^2 = d\alpha^2 + \sin^2\alpha  d\tilde{\phi}^2  + \frac{1}{4} \left(\frac{dr^2}{ \cos\alpha^2}  +  \frac{r^2 \cos^2 \alpha} { r^2+ \cos^4 \alpha } (d\theta^2 +\sin^2 \theta d\phi^2 )\right) \ . 
\ee
This internal metric when augmented by the untouched $AdS_5$ space-time and supplemented by the dilaton and NS and RR fluxes provides the solution of \cite{Sfetsos:2010uq} preserving ${\cal N}=2$ supersymmetry. 

The Buscher procedure gives the T-duality rules for world-sheet derivatives which read
\be
l_{+}^a \to -( M^{-1})^{ba } \partial_+ v_b  \ , \quad  l_{-}^a \to  ( M^{-1})^{ab } \partial_+ v_b  \ .
\ee
  Upon making the above substitution into eqs.~\eqref{eq:a0} and \eqref{eq:a2}  the Lax connection of \eqref{eq:LaxCoset} becomes T-dualised to a Lax connection encoding the dynamics of the T-dual theory with Lagrangian eq.~\eqref{eq:Lagcosetdual}.   
  
    Since in the bosonic theory the $AdS_5$ is coupled to the $S^5$ only via the Virasoro constraints (which are preserved in the T-dualisation),  this guarantees the classical integrability (in the bosonic sector) of the  $SU(2)_L$ non-Abelian T-dual of $AdS_5 \times S^5$.   The full  $\frak{psu}(2,2 | 4)$ coset is treated in  \cite{Borsato:2017qsx}.


\begin{thebibliography}{99}



{\small


\bibitem{Maldacena:1997re} 
  J.~M.~Maldacena,
``The Large N limit of superconformal field theories and supergravity,''
  Int.\ J.\ Theor.\ Phys.\  {\bf 38}, 1113 (1999)
  [Adv.\ Theor.\ Math.\ Phys.\  {\bf 2}, 231 (1998)]
  \arxivlink{hep-th/9711200}.


\bibitem{Beisert:2010jr} 
For a not so  recent review see
  N.~Beisert {\it et al.},
``Review of AdS/CFT Integrability: An Overview,''
  Lett.\ Math.\ Phys.\  {\bf 99}, 3 (2012)
  [\arxivlink{1012.3982}].
  
  
\bibitem{Bena:2003wd}
  I.~Bena, J.~Polchinski and R.~Roiban,
   ``Hidden symmetries of the $AdS_5 \times S^5$ superstring,''
  Phys.\ Rev.\ D {\bf 69} (2004) 046002
  \arxivlink{hep-th/0305116}.
  
    
  \bibitem{Leigh:1995ep} 
 S.~Frolov,
``Lax pair for strings in Lunin-Maldacena background,''
  JHEP {\bf 0505}, 069 (2005)
  \arxivlink{hep-th/0503201}.
  
  \bibitem{Roiban:2003dw} 
  R.~Roiban,
``On spin chains and field theories,''
  JHEP {\bf 0409}, 023 (2004)
  \arxivlink{hep-th/0312218}.
  
  \bibitem{Frolov:2005ty} 
  S.~A.~Frolov, R.~Roiban and A.~A.~Tseytlin,
``Gauge-string duality for superconformal deformations of N=4 super Yang-Mills theory,''
  JHEP {\bf 0507}, 045 (2005)
  \arxivlink{hep-th/0503192}.
  
  \bibitem{Berenstein:2004ys} 
  D.~Berenstein and S.~A.~Cherkis,
``Deformations of N=4 SYM and integrable spin chain models,''
  Nucl.\ Phys.\ B {\bf 702}, 49 (2004)
  \arxivlink{hep-th/0405215}.


\bibitem{Giataganas:2013dha}
  D.~Giataganas, L.~A.~Pando Zayas and K.~Zoubos,
``On Marginal Deformations and Non-Integrability,''
  JHEP {\bf 1401} (2014) 129
  [\arxivlink{1311.3241}].


\bibitem{Klebanov:1998hh} 
  I.~R.~Klebanov and E.~Witten,
``Superconformal field theory on three-branes at a Calabi-Yau singularity,''
  Nucl.\ Phys.\ B {\bf 536}, 199 (1998)
  \arxivlink{hep-th/9807080}.
  
\bibitem{Basu:2011di}
  P.~Basu and L.~A.~Pando Zayas,
``Chaos rules out integrability of strings on AdS$_5 \times T^{1,1}$,''
  Phys.\ Lett.\ B {\bf 700} (2011) 243
  [\arxivlink{1103.4107}].

  
  
\bibitem{Sfetsos:2013wia}
  K.~Sfetsos,
``Integrable interpolations: From exact CFTs to non-Abelian T-duals,''
  Nucl.\ Phys.\ B {\bf 880} (2014) 225
  [\arxivlink{1312.4560}].
\bibitem{Hollowood:2014qma}
  T.~J.~Hollowood, J.~L.~Miramontes and D.~M.~Schmidtt,
``An Integrable Deformation of the $AdS_5 \times S^5$ Superstring,''
  J.\ Phys.\ A {\bf 47} (2014) no.49,  495402
  [\arxivlink{1409.1538}].
  
  
\bibitem{Klimcik:2002zj}
  C.~Klimcik,
``Yang-Baxter sigma models and dS/AdS T duality,''
  JHEP {\bf 0212} (2002) 051
  \arxivlink{hep-th/0210095}.
  
\bibitem{Delduc:2013qra}
  F.~Delduc, M.~Magro and B.~Vicedo,
``An integrable deformation of the $AdS_5 x S^5$ superstring action,''
  Phys.\ Rev.\ Lett.\  {\bf 112} (2014) no.5,  051601
  [\arxivlink{1309.5850}].

\bibitem{Sfetsos:2014cea}
  K.~Sfetsos and D.~C.~Thompson,
``Spacetimes for $\lambda$-deformations,''
  JHEP {\bf 1412} (2014) 164
  [\arxivlink{1410.1886}].
  
\bibitem{Arutyunov:2015mqj}
  G.~Arutyunov, S.~Frolov, B.~Hoare, R.~Roiban and A.~A.~Tseytlin,
``Scale invariance of the $\eta$-deformed $AdS_5\times S^5$ superstring, T-duality and modified type II equations,''
  Nucl.\ Phys.\ B {\bf 903} (2016) 262
  [\arxivlink{1511.05795}].
\bibitem{Borsato:2016zcf}
  R.~Borsato, A.~A.~Tseytlin and L.~Wulff,
``Supergravity background of $\lambda$-deformed model for AdS$_2 \times$  S$^2$ supercoset,''
  Nucl.\ Phys.\ B {\bf 905} (2016) 264
  [\arxivlink{1601.08192}].
  
\bibitem{Chervonyi:2016ajp}
  Y.~Chervonyi and O.~Lunin,
``Supergravity background of the $\lambda$-deformed AdS$_3 \times$ S$^3$ supercoset,''
  Nucl.\ Phys.\ B {\bf 910} (2016) 685
  [\arxivlink{1606.00394}].
  
  
\bibitem{Borsato:2016ose}
  R.~Borsato and L.~Wulff,
``Target space supergeometry of $\eta$ and $\lambda$-deformed strings,''
  JHEP {\bf 1610} (2016) 045
  [\arxivlink{1608.03570}].








\bibitem{Zayas:2010fs}
  L.~A.~Pando Zayas and C.~A.~Terrero-Escalante,
``Chaos in the Gauge / Gravity Correspondence,''
  JHEP {\bf 1009} (2010) 094
  [\arxivlink{1007.0277}].

\bibitem{Basu:2011dg}
  P.~Basu, D.~Das and A.~Ghosh,
``Integrability Lost,''
  Phys.\ Lett.\ B {\bf 699} (2011) 388
  [\arxivlink{1103.4101}].


\bibitem{Basu:2011fw}
  P.~Basu and L.~A.~Pando Zayas,
``Analytic Non-integrability in String Theory,''
  Phys.\ Rev.\ D {\bf 84} (2011) 046006
  [\arxivlink{1105.2540}].
 
 \bibitem{Basu:2012ae} 
  P.~Basu, D.~Das, A.~Ghosh and L.~A.~Pando Zayas,
  JHEP {\bf 1205}, 077 (2012)
  doi:10.1007/JHEP05(2012)077
  [arXiv:1201.5634 [hep-th]].



\bibitem{Stepanchuk:2012xi}
  A.~Stepanchuk and A.~A.~Tseytlin,
``On (non)integrability of classical strings in p-brane backgrounds,''
  J.\ Phys.\ A {\bf 46} (2013) 125401
  [\arxivlink{1211.3727}].
  
  

\bibitem{Chervonyi:2013eja}
  Y.~Chervonyi and O.~Lunin,
``(Non)-Integrability of Geodesics in D-brane Backgrounds,''
  JHEP {\bf 1402} (2014) 061
  [\arxivlink{1311.1521}].
  
  
\bibitem{Giataganas:2014hma}
  D.~Giataganas and K.~Sfetsos,
``Non-integrability in non-relativistic theories,''
  JHEP {\bf 1406} (2014) 018
  [\arxivlink{1403.2703}].
  
\bibitem{Asano:2015eha}
  Y.~Asano, D.~Kawai and K.~Yoshida,
``Chaos in the BMN matrix model,''
  JHEP {\bf 1506} (2015) 191
  [\arxivlink{1503.04594}].

  
\bibitem{Asano:2015qwa}
  Y.~Asano, D.~Kawai, H.~Kyono and K.~Yoshida,
``Chaotic strings in a near Penrose limit of AdS$_{5} \times$ T$^{1,1}$,''
  JHEP {\bf 1508} (2015) 060
  [\arxivlink{1505.07583}].
  
\bibitem{Ishii:2016rlk}
  T.~Ishii, K.~Murata and K.~Yoshida,
``Fate of chaotic strings in a confining geometry,''
  Phys.\ Rev.\ D {\bf 95} (2017) no.6,  066019
  [\arxivlink{1610.05833}].
  
  
  \bibitem{Panigrahi:2016zny} 
  K.~L.~Panigrahi and M.~Samal,
  ``Chaos in classical string dynamics in $\hat{\gamma}$ deformed $AdS_5 \times T^{1,1}$,''
  Phys.\ Lett.\ B {\bf 761}, 475 (2016)
  doi:10.1016/j.physletb.2016.08.021
  [\arxivlink{1605.05638}].
  
\bibitem{Basu:2016zkr} 
  P.~Basu, P.~Chaturvedi and P.~Samantray,
  ``Chaotic dynamics of strings in charged black hole backgrounds,''
  Phys.\ Rev.\ D {\bf 95}, no. 6, 066014 (2017)
  doi:10.1103/PhysRevD.95.066014 
    [\arxivlink{1607.04466}].


\bibitem{Giataganas:2017guj} 
  D.~Giataganas and K.~Zoubos,
 ``Non-integrability and Chaos with Unquenched Flavor,''
  JHEP {\bf 1710}, 042 (2017)
  doi:10.1007/JHEP10(2017)042 
      [\arxivlink{1707.04033}].

\bibitem{Nunez:2018ags}
  C.~Nunez, J.~M.~Penin, D.~Roychowdhury and J.~Van Gorsel,
   ``The non-Integrability of Strings in Massive Type IIA and their Holographic duals,''
  arXiv:1802.04269 [hep-th].
        [\arxivlink{1802.04269}].

\bibitem{Roychowdhury:2017vdo}
  D.~Roychowdhury,
``Analytic integrability for strings on $ \eta $ and $ \lambda $ deformed backgrounds,''
  JHEP {\bf 1710} (2017) 056
  [\arxivlink{1707.07172}].

  
  
\bibitem{Kovacic85}
J.~J. Kovacic, ``An algorithm for solving second order linear homogeneous
  differential equations,'' {\em J. Symb. Comp} {\bf 2} (1986) 3--43

 
\bibitem{Gaiotto:2009we}
  D.~Gaiotto,
``N=2 dualities,''
  JHEP {\bf 1208} (2012) 034
  [\arxivlink{0904.2715}].


\bibitem{Gaiotto:2009gz}
  D.~Gaiotto and J.~Maldacena,
``The Gravity duals of N=2 superconformal field theories,''
  JHEP {\bf 1210} (2012) 189
  [\arxivlink{0904.4466}].
  

\bibitem{Sfetsos:2010uq} 
  K.~Sfetsos and D.~C.~Thompson,
``On non-abelian T-dual geometries with Ramond fluxes,''
  Nucl.\ Phys.\ B {\bf 846}, 21 (2011)
  [\arxivlink{1012.1320}].
  
  
\bibitem{Maldacena:2000mw} 
  J.~M.~Maldacena and C.~Nunez,
``Supergravity description of field theories on curved manifolds and a no go theorem,''
  Int.\ J.\ Mod.\ Phys.\ A {\bf 16}, 822 (2001)
  \arxivlink{hep-th/0007018}.
  
\bibitem{ReidEdwards:2010qs} 
  R.~A.~Reid-Edwards and B.~Stefanski, jr.,
``On Type IIA geometries dual to N = 2 SCFTs,''
  Nucl.\ Phys.\ B {\bf 849}, 549 (2011)
  [\arxivlink{1011.0216}].
\bibitem{Aharony:2012tz} 
  O.~Aharony, L.~Berdichevsky and M.~Berkooz,
``4d N=2 superconformal linear quivers with type IIA duals,''
  JHEP {\bf 1208}, 131 (2012)
  [\arxivlink{1206.5916}].
  
\bibitem{Macpherson:2015tka} 
  N.~T.~Macpherson, C.~Nunez, D.~C.~Thompson and S.~Zacarias,
``Holographic Flows in non-Abelian T-dual Geometries,''
  JHEP {\bf 1511}, 212 (2015)
  [\arxivlink{1509.04286}].
  
  
  \bibitem{Lozano:1995jx}
  Y.~Lozano,
   ``NonAbelian duality and canonical transformations,''
  Phys.\ Lett.\ B {\bf 355} (1995) 165
  \arxivlink{hep-th/9503045}.
\bibitem{Borsato:2016pas}
  R.~Borsato and L.~Wulff,
``Integrable Deformations of $T$-Dual $\sigma$ Models,''
  Phys.\ Rev.\ Lett.\  {\bf 117} (2016) no.25,  251602
  [\arxivlink{1609.09834}].
\bibitem{Borsato:2017qsx}
  R.~Borsato and L.~Wulff,
``On non-abelian T-duality and deformations of supercoset string sigma-models,''
  JHEP {\bf 1710} (2017) 024
  [\arxivlink{1706.10169}].
  
\bibitem{Kawaguchi:2014qwa}
  I.~Kawaguchi, T.~Matsumoto and K.~Yoshida,
   ``Jordanian deformations of the $AdS_5 x S^5$ superstring,''
  JHEP {\bf 1404} (2014) 153
  doi:10.1007/JHEP04(2014)153
    [\arxivlink{1401.4855}]. 
  
\bibitem{Vicedo:2011zz}
  B.~Vicedo,
``The method of finite-gap integration in classical and semi-classical string theory,''
  J.\ Phys.\ A {\bf 44} (2011) 124002
  [\arxivlink{0810.3402}].
  
  \bibitem{delaOssa:1992vci}
  X.~C.~de la Ossa and F.~Quevedo,
``Duality symmetries from nonAbelian isometries in string theory,''
  Nucl.\ Phys.\ B {\bf 403} (1993) 377
  \arxivlink{hep-th/9210021}.
  
  
\bibitem{Itsios:2017nou} 
  G.~Itsios, H.~Nastase, C.~Nunez, K.~Sfetsos and S.~Zacarias,
``Penrose limits of Abelian and non-Abelian T-duals of $AdS_5\times S^5$ and their field theory duals,''
  JHEP {\bf 1801}, 071 (2018)
  [\arxivlink{1711.09911}].
    
\bibitem{Lozano:2016kum} 
  Y.~Lozano and C.~Nunez,
``Field theory aspects of non-Abelian T-duality and $ \mathcal{N}  =$ 2 linear quivers,''
  JHEP {\bf 1605}, 107 (2016)
  [\arxivlink{1603.04440}].
  

\bibitem{Singer1}
M. van der Put and M. F. Singer, {\em Galois Theory of Linear Differential Equation}  Springer-Verlag, 2003
  
\bibitem{Singer2}
M. F. Singer, {\em  Introduction to the Galois Theory of Linear Differential Equations Algebraic Theory of Differential Equations}, M.A.H. MacCallum and A.V. Mikhalov, eds., London Mathematical Society Lecture Note Series (no. 357), Cambridge University Press, 2009, 1-82.

\bibitem{Ruiz:1999}
J.~J.~M. Ruiz, {\em Differential Galois theory and non-integrability of
  Hamiltonian systems}  Birkhauser, 1999
  
  \bibitem{MoralesRuizRamis2001}
J.~J. Morales-Ruiz and J.-P. Ramis, ``Galoisian obstructions to integrability
  of hamiltonian systems,'' {\em Methods and Applications of Analysis} {\bf 8}
  (March, 2001) 33 -- 96 
 


  \bibitem{ziglin1982}
S.~L. Ziglin, ``Branching of solutions and nonexistence of first integrals in
  hamiltonian mechanics. i,'' {\em Functional Analysis and Its Applications}
  {\bf 16} (1982) 181--189
  
 
  \bibitem{Morales1994140}
J.~Morales and C.~Simo, ``Picard-Vessiot theory and Ziglin's theorem,'' {\em
  Journal of Differential Equations} {\bf 107} (1994), no.~1, 140 -- 162
  
  \bibitem{MoralesRuiz2007845}
J.~J. Morales-Ruiz, J.-P. Ramis, and C.~Simo, ``Integrability of hamiltonian
  systems and differential Galois groups of higher variational equations,''
 {\em Annales Scientifiques de l'\'Ecole Normale Sup\'erieure} {\bf 40}
  (2007), no.~6, 845 -- 884
%
%
%
%
%

  
  
   
  
  










 
\bibitem{Lozano:2017ole}
  Y.~Lozano, C.~Nunez and S.~Zacarias,
    ``BMN Vacua, Superstars and Non-Abelian T-duality,''
  JHEP {\bf 1709} (2017) 000
  [\arxivlink{1703.00417}].  
   
  
    
    \bibitem{Ott:2002book} 
  E.~Ott,
  \emph{Chaos in Dynamical Systems}
  Cambridge: Cambridge University Press

\bibitem{Arutyunov:2009ga}
  G.~Arutyunov and S.~Frolov,
   ``Foundations of the $AdS_5 x S^5$ Superstring. Part I,''
  J.\ Phys.\ A {\bf 42} (2009) 254003
  [\arxivlink{0901.4937}].
%
%
%
%
  
\bibitem{Assel:2011xz} 
  B.~Assel, C.~Bachas, J.~Estes and J.~Gomis,
  JHEP {\bf 1108}, 087 (2011)
  doi:10.1007/JHEP08(2011)087
  [arXiv:1106.4253 [hep-th]].
  
\bibitem{DHoker:2017zwj} 
  E.~D'Hoker, M.~Gutperle and C.~F.~Uhlemann,
  JHEP {\bf 1711}, 200 (2017)
  doi:10.1007/JHEP11(2017)200
  [arXiv:1706.00433 [hep-th]].
  
   
\bibitem{Lozano:2016wrs} 
  Y.~Lozano, N.~T.~Macpherson, J.~Montero and C.~Nunez,
  JHEP {\bf 1611}, 133 (2016)
  doi:10.1007/JHEP11(2016)133
  [arXiv:1609.09061 [hep-th]].
  
  
\bibitem{Lozano:2012au} 
  Y.~Lozano, E.O Colgain, D.~Rodriguez-Gomez and K.~Sfetsos,
  Phys.\ Rev.\ Lett.\  {\bf 110}, no. 23, 231601 (2013)
  doi:10.1103/PhysRevLett.110.231601
  [arXiv:1212.1043 [hep-th]].

  
  
\bibitem{Gadde:2012rv}
 A.~Gadde, P.~Liendo, L.~Rastelli and W.~Yan,
 JHEP {\bf 1308}, 015 (2013)
 doi:10.1007/JHEP08(2013)015
 [arXiv:1211.0271 [hep-th]].


\bibitem{Pomoni:2013poa}
 E.~Pomoni,
 Nucl.\ Phys.\ B {\bf 893}, 21 (2015)
 doi:10.1016/j.nuclphysb.2015.01.006
 [arXiv:1310.5709 [hep-th]].
  
}\end{thebibliography}
\end{document}